\documentclass[english,prd,english,notitlepage,nofootinbib]{revtex4}
\usepackage[latin9]{inputenc}
\usepackage{xcolor}
\usepackage{pdfcolmk}
\usepackage{babel}
\usepackage{amsmath}
\usepackage{amssymb}
\usepackage{graphicx}
\usepackage{esint}
\PassOptionsToPackage{normalem}{ulem}
\usepackage{ulem}
\usepackage[unicode=true,pdfusetitle,
 bookmarks=true,bookmarksnumbered=false,bookmarksopen=false,
 breaklinks=false,pdfborder={0 0 1},backref=false,colorlinks=false]
 {hyperref}

\makeatletter

\providecommand{\tabularnewline}{\\}
\providecolor{lyxadded}{rgb}{0,0,1}
\providecolor{lyxdeleted}{rgb}{1,0,0}

\DeclareRobustCommand{\lyxdeleted}[3]{{\texorpdfstring{\color{lyxdeleted}\sout{#3}}{}}}

\@ifundefined{textcolor}{}
{%
 \definecolor{BLACK}{gray}{0}
 \definecolor{WHITE}{gray}{1}
 \definecolor{RED}{rgb}{1,0,0}
 \definecolor{GREEN}{rgb}{0,1,0}
 \definecolor{BLUE}{rgb}{0,0,1}
 \definecolor{CYAN}{cmyk}{1,0,0,0}
 \definecolor{MAGENTA}{cmyk}{0,1,0,0}
 \definecolor{YELLOW}{cmyk}{0,0,1,0}
}

\usepackage{babel}

\providecommand{\tabularnewline}{\\}

\@ifundefined{textcolor}{}{%
 \definecolor{BLACK}{gray}{0}
 \definecolor{WHITE}{gray}{1}
 \definecolor{RED}{rgb}{1,0,0}
 \definecolor{GREEN}{rgb}{0,1,0}
 \definecolor{BLUE}{rgb}{0,0,1}
 \definecolor{CYAN}{cmyk}{1,0,0,0}
 \definecolor{MAGENTA}{cmyk}{0,1,0,0}
 \definecolor{YELLOW}{cmyk}{0,0,1,0}
}

\usepackage{babel}

\makeatother

\begin{document}

\title{Twist-three contributions to Charged Current Deeply Virtual Meson
production (CCDVMP)}

\title{$\rho$- vs . $\pi$-production in Charged Current Deeply Virtual
Meson production (CCDVMP)}

\title{What studies DVMP: nucleon GPDs or meson DAs ?}

\title{Model-independent tests of meson DAs from DVMP}

\title{Generalized Parton Distributions from charged current meson production}

\author{Marat~Siddikov, Iván~Schmidt}

\address{Departamento de Física, Universidad Técnica Federico Santa María,\\
 y Centro Científico - Tecnológico de Valparaíso, Casilla 110-V, Valparaíso,
Chile}
\begin{abstract}
In this paper we prove that the simultaneous study of both $\rho$- and $\pi$-meson
production by charged currents in Bjorken kinematics allows for a very
clean extraction of the leading twist Generalized Parton Distributions
of the target, with inherent control of the contribution of higher-twist
corrections. Also, it might provide target-independent constraints
on the distribution amplitudes of the produced mesons. We expect that
such processes might be studied either in neutrino-induced or in electron-induced
processes. According to our numerical estimates, the cross-sections
of these processes are within the reach of JLab and EIC experiments.
\end{abstract}

\pacs{13.15.+g,13.85.-t}

\keywords{Single pion production, generalized parton distributions, electon-hadron
interactions.}
\maketitle

\section{Introduction}

The structure of the hadrons remains up to now a challenging puzzle,
which attracts a lot of attention from both theoretical and experimental
viewpoints. Nowadays, this structure is parametrized in terms of the
so-called generalized parton distributions (GPDs), which are directly
related to amplitudes of physical processes in Bjorken kinematics~\cite{Ji:1998xh,Collins:1998be}.
The early analyses of GPDs were mostly based on experimental data
on deeply virtual Compton scattering (DVCS)~\cite{Dupre:2017hfs}
and deeply virtual meson production (DVMP)~\cite{Mueller:1998fv,Ji:1996nm,Ji:1998pc,Radyushkin:1996nd,Radyushkin:1997ki,Radyushkin:2000uy,Collins:1996fb,Brodsky:1994kf,Goeke:2001tz,Diehl:2000xz,Belitsky:2001ns,Diehl:2003ny,Belitsky:2005qn,Kubarovsky:2011zz},
yet very soon it was realized that in view of the rich structure of
GPDs, the poorly known wave functions of the produced mesons, as well as the sizable
higher twist contributions~\cite{Kubarovsky:2011zz,Ahmad:2008hp,Goloskokov:2009ia,Goloskokov:2011rd,Goldstein:2012az},
additional channels are needed. Since the amplitudes of physical processes
typically include contributions of GPDs of several flavors and helicity
states (sometimes convoluted with distribution amplitudes of other
hadrons), the GPDs could be extracted only from self-consistent global
fits of all available experimental data. Currently the list of processes
which might be used for the extraction of GPDs include: $\rho$-meson
photoproduction~\cite{Anikin:2009bf,Diehl:1998pd,Mankiewicz:1998kg,Mankiewicz:1999tt,Boussarie:2017umz},
timelike Compton Scattering~\cite{Berger:2001xd,Pire:2008ea,Boer:2015fwa},
exclusive pion- or photon-induced lepton pair production~\cite{Muller:2012yq,Sawada:2016mao},
heavy charmonia photoproduction~\cite{Ivanov:2004vd,Ivanov:2015hca}
(for gluon GPDs), as well as a few other channels~\cite{Kofler:2014yka,Accardi:2012qut}.
Hopefully the forthcoming experimental data from upgraded JLab~\cite{Kubarovsky:2011zz},
COMPASS~\cite{Gautheron:2010wva,Kouznetsov:2016vvo,Ferrero:2012ega,Sandacz:2016kwh,Sandacz:2017ctv,Silva:2013dta}
and J-PARC~\cite{Sawada:2016mao,Kroll:2016kvd}, will enrich and enhance
the early data from HERA and 6 GeV JLab experiments, as well as improve
our understanding of the GPDs of the proton~\cite{Kroll:2019wug,Kroll:2018uvl,Anikin:2017fwu,Kroll:2017hym,Airapetian:2017vit,Kroll:2016aop,Favart:2015umi,Kumericki:2017gdc,Kumericki:2016ela,Duplancic:2018bum,Duplancic:2016bge,Pire:2017yge}.

Some of the experimentally studied channels suffer from well-understood
theoretical complications. For example, as was found recently from theoretical analysis of pion DVMP~\cite{Defurne:2016eiy},
the dominant contribution in JLab kinematics (and possibly
at the planned Electron Ion Collider~\cite{Accardi:2012qut}) stems from
transversely polarized virtual photons, which implies dominance
of twist-three effects. A careful Rosenbluth separation might
help to single out contributions of the longitudinal photons. However,
even in this case the longitudinal cross-sections might still include
various other sources of higher-twist contributions~\cite{Anikin:2009bf}.
Recently it was suggested that a test of the $Q^{2}$-dependence~\cite{THorn}
might be used to check if the description of $\sigma_{L}$ based on
the leading twist collinear factorization predictions is correct~.
However, this method might give reliable estimates provided data
at sufficiently large $Q^{2}$ are available. Another challenge for the
present analyses of DVMP is unknown distribution amplitudes (DAs) of
mesons. While it is expected that the DA should be close to their asymptotic
form~\cite{Fu:2016yzx,Bali:2017gfr}, due to the structure of the
DVMP amplitude in the next-to-leading order, the currently admitted
deviations of DA from the asymptotic form might lead to sizable (up to
50 per cent) deviations of the cross-section~\cite{Diehl:2003ny,Ivanov:2004vd,Ivanov:2004zv,Ivanov:2015hca,Diehl:2007hd}.

In this paper we propose a novel method which allows to extract GPDs, as
well as have a simultaneous control of the twist-three effects and
the uncertainty in the distribution amplitudes. Our approach is based
on comparison of $\rho$- and $\pi$-meson production cross-sections
in charged current processes. In fact, the feasibility of using charged current
processes for study of GPDs was demonstrated in~\cite{Pire:2015iza,Pire:2015vxa,Pire:2016jtr,Pire:2017lfj,Pire:2017tvv,Siddikov:2016zmt},
with possible application either to neutrino-induced~\cite{Drakoulakos:2004gn}
or to electron-induced channels~\footnote{The feasibility to study experimentally the charged currents in JLAB
kinematics was demonstrated earlier in~\cite{Androic:2013rhu}.
It is expected that after the upgrade, higher instant luminosities
up to $\mathcal{L}=10^{38}{\rm cm}^{-2}\cdot s^{-1}$ will be achieved~\cite{Alcorn:2004sb},
which implies that the DVMP cross-section could be measured with reasonable
statistics. The neutrino kinematics might be reconstructed using missing
mass techniques.}. These processes have a small contamination by twist-3 effects~\cite{Kopeliovich:2014pea},
and on an unpolarized target they get their dominant contribution from the GPDs
$H_{u},\,H_{d}$. Due to the $V-A$ structure of the hadronic current,
in leading twist the CCDVMP cross-sections of longitudinally polarized
$\rho$- mesons and pions are sensitive to exactly the same set of
GPDs and thus allow for a variety of consistency checks.

In this paper we will focus on the main contribution to the production of longitudinally
polarized $\rho_{L}^{\pm}$-mesons, which can be evaluated in the
collinear factorization framework~\cite{Anikin:2009bf,Diehl:1998pd,Mankiewicz:1998kg,Mankiewicz:1999tt,Boussarie:2017umz,Kopeliovich:2013ae}
and gives the dominant contribution in the Bjorken limit. Due to the $V-A$
structure of the hadronic current, the cross-sections of the $\rho_{L}^{\pm}$-
and $\pi^{\pm}$-meson production are controlled by the same combination
of GPDs, so any differences between the two cross-sections comes
only from the meson wave functions or higher twist effects.  In 
leading order, the dependence on meson distribution amplitudes contributes
only as a multiplicative prefactor, so the ratio of the cross-sections
\begin{equation}
R_{\rho/\pi}\left(x_{B},\,Q^{2}\right)=\frac{d\sigma_{W^{\pm}p\to\rho^{\pm}p}}{d\sigma_{W^{\pm}p\to\pi^{\pm}p}}\approx{\rm const},\label{eq:RRatio}
\end{equation}
does not depend on the GPDs of the target. In this approximation the ratio
is the same for both proton and neutron targets ($W^{\pm}n\to M^{\pm}n$
subprocess), and for this reason it might be studied on nuclear targets
instead of protons. In phenomenological models it is frequently speculated
that the leading twist distribution amplitudes of pion and $\rho$-meson
are close to their asymptotic form, so the ratio should be close to $\left(f_{\rho}/f_{\pi}\right)^{2}$,
where $f_{\rho},\,f_{\pi}$ are the corresponding decay constants
of $\rho$ and $\pi$ mesons. The deviations from this value are due
to deviations from the asymptotic form of distribution amplitudes, and next-to-leading
order and higher-twist corrections. Each of such corrections has a
characteristic behavior in the $\left(x_{B},\,Q^{2}\right)$ variables, which
can be used to clearly distinguish its origin. For this
reason we believe that the ratio~(\ref{eq:RRatio}) is a sensitive
probe of the leading twist contribution dominance,
as well as of tests of the meson distribution amplitudes. In the
following sections we will discuss in detail how the value of this
ratio changes when NLO corrections and higher twist effects are taken
into account. For the sake of brevity and conciseness, in this paper
we do not consider other processes, where flavor multiplet partners
of pions and protons are produced and which could also be used to
test other flavor combinations of pion and $\rho$-meson distribution
amplitudes. 

The paper is organized as follows. In Section~\ref{sec:DVMP_Xsec}
we discuss the framework used for the evaluation of meson production,
taking into account NLO and some of the higher twist-corrections.
In Section~\ref{subsec:DAs} we define amplitudes of $\rho$-mesons
and pions and discuss their parameterization. In Section~\ref{sec:DVMP_Xsec}
we present expressions for the cross-sections of the CCDVMP process
in the leading twist. In Section~\ref{sec:Tw3} we discuss the contribution
of twist-three corrections to the cross-section. Finally, in Section~\ref{sec:Results}
we present numerical results and draw conclusions. 

\section{The CCDVMP process}

\subsection{Meson distribution amplitudes}

\label{subsec:DAs}For the sake of completeness we would like to start the
discussion with explicit definitions of the distribution amplitudes
of the pion and $\rho$-meson. We will consider only the two-parton
DAs. For the pion case, the corresponding DAs are defined as~\cite{Ball:2006wn,Kopeliovich:2011rv}
\begin{eqnarray}
\left\langle 0\left|\bar{\psi}\left(y\right)\gamma_{\mu}\gamma_{5}\psi\left(x\right)\right|\pi(q)\right\rangle  & = & if_{\pi}\int_{0}^{1}d\alpha\,e^{i(\alpha p\cdot y+\bar{\alpha}p\cdot x)}\times\nonumber \\
 & \times & \left(p_{\mu}\phi_{2;\pi}(\alpha)+\frac{1}{2}\frac{z_{\mu}}{(p\cdot z)}\psi_{4;\pi}(\alpha)\right),\label{eq:piWF-mu5}
\end{eqnarray}

\begin{eqnarray}
\left\langle 0\left|\bar{\psi}\left(y\right)\gamma_{5}\psi\left(x\right)\right|\pi(q)\right\rangle  & = & -if_{\pi}\frac{m_{\pi}^{2}}{m_{u}+m_{d}}\int_{0}^{1}d\alpha\,e^{i(\alpha p\cdot y+\bar{\alpha}p\cdot x)}\phi_{3;\pi}^{(p)}(\alpha),\label{eq:piWF-5}
\end{eqnarray}

\begin{eqnarray}
\left\langle 0\left|\bar{\psi}\left(y\right)\sigma_{\mu\nu}\gamma_{5}\psi\left(x\right)\right|\pi(q)\right\rangle  & = & -\frac{i}{3}f_{\pi}\frac{m_{\pi}^{2}}{m_{u}+m_{d}}\int_{0}^{1}d\alpha\,e^{i(\alpha p\cdot y+\bar{\alpha}p\cdot x)}\times\nonumber \\
 & \times & \frac{1}{p\cdot z}\left(p_{\mu}z_{\nu}-p_{\nu}z_{\mu}\right)\phi_{3;\pi}^{(\sigma)}(\alpha).\label{eq:piWF-munu}
\end{eqnarray}
where $q$ is the momentum of the pion, $z\equiv x-y$ is the light-cone
separation of the quarks, $p$ is the light-cone vector bound by $p^{2}=0$,~~~$p\cdot z=1$;
$f_{\pi}$ is the pion decay constant, $m_{\pi}$ is the pion mass, and
$m_{u}$ and $m_{d}$ are masses of the $u$ and $d$ quarks respectively.
In what follows we will focus on the twist-2 and twist-3 DAs $\phi_{2;\pi}$,
$\phi_{3;\pi}^{(p)}$ and $\phi_{3;\pi}^{(\sigma)}$. Similarly, for
the case of $\rho$-meson, the distribution amplitudes are defined
as~\cite{Ball:1998sk}
\begin{eqnarray}
\left\langle 0\left|\bar{\psi}\left(y\right)\gamma_{\mu}\psi\left(x\right)\right|\rho(q)\right\rangle  & = & f_{\rho}m_{\rho}\int_{0}^{1}d\alpha\,e^{i(0.5-\alpha)p\cdot z}\times\nonumber \\
 & \times & \left(p_{\mu}\frac{e^{(\lambda)}\cdot z}{p\cdot z}\phi_{2,\rho}^{(||)}(\alpha)+e_{\mu}^{(\lambda=\perp)}g_{\perp}^{(v)}(\alpha)-\frac{m_{\rho}^{2}}{2}z_{\mu}\frac{e^{(\lambda)}\cdot z}{(p\cdot z)^{2}}g_{3}(\alpha)\right),\label{eq:AWF-mu5-alt}
\end{eqnarray}

\begin{eqnarray}
\left\langle 0\left|\bar{\psi}\left(y\right)\gamma_{\mu}\gamma_{5}\psi\left(x\right)\right|\rho(q)\right\rangle  & = & \frac{1}{2}\left(f_{\rho}-f_{\rho}^{T}\frac{m_{u}+m_{d}}{m_{\rho}}\right)m_{\rho}\epsilon_{\mu\nu\rho\sigma}e_{\nu}^{(\lambda)}p_{\rho}z_{\sigma}\int_{0}^{1}d\alpha\,e^{i(0.5-\alpha)p\cdot z}g_{\perp}^{(a)}(\alpha)\label{eq:AWF-mu-alt}
\end{eqnarray}

\begin{eqnarray}
\left\langle 0\left|\bar{\psi}\left(y\right)\sigma_{\mu\nu}\psi\left(x\right)\right|\rho(q)\right\rangle  & = & if_{\rho}^{T}\int_{0}^{1}d\alpha\,e^{i(0.5-\alpha)p\cdot z}\left(\left(e_{\mu}^{(\lambda=\perp)}p_{\nu}-e_{\nu}^{(\lambda=\perp)}p_{\mu}\right)\Phi_{\perp}(\alpha)\right.+\nonumber \\
 & + & \frac{e^{(\lambda)}\cdot z}{(p\cdot z)^{2}}m_{\rho}^{2}\left(p_{\mu}z_{\nu}-p_{\nu}z_{\mu}\right)h_{||}^{(t)}(\alpha)\label{eq:AWF-munu-alt}\\
 & + & \left.\frac{1}{2}\left(e_{\mu}^{(\lambda)}z_{\nu}-e_{\nu}^{(\lambda)}z_{\mu}\right)\frac{m_{\rho}^{2}}{p\cdot z}\,h_{3}(\alpha)\right),\nonumber 
\end{eqnarray}

\begin{eqnarray}
\left\langle 0\left|\bar{\psi}\left(y\right)\psi\left(x\right)\right|\rho(q)\right\rangle  & =-i & \left(f_{\rho}^{T}-f_{\rho}\frac{m_{u}+m_{d}}{m_{\rho}}\right)e^{(\lambda)}\cdot n\int_{0}^{1}d\alpha\,e^{i(0.5-\alpha)p\cdot z}h_{||}^{(s)}(\alpha).\label{eq:AWF-5-alt}
\end{eqnarray}
where $f_{\rho}$ and $f_{\rho}^{T}$ are the so-called vector and
tensor decay constants, and $m_{\rho}$ is the $\rho$-meson mass.
In what follows we will focus on the contribution for the longitudinal
mesons (for which factorization has been proven) and consider only
the contributions up to twist 3, $\Phi_{||},\,h_{||}^{(s)}$ and $h_{||}^{(t)}$.
As we can see, the pion and $\rho$-meson distribution amplitudes
differ from each other only by an additional $\gamma_{5}$ in the quark-antiquark
operator (modulo some trivial numerical prefactor). In the next section
we will show that due to this property, the CCDVMP amplitudes of $\rho$-meson
and pion are related to each other by a mere substitution of meson
DAs, 
\begin{equation}
f_{\pi}\phi_{2;\pi}(\alpha)\leftrightarrow f_{\rho}\phi_{2,\rho}^{(||)}(\alpha),\label{eq:tw2_sub}
\end{equation}
\begin{equation}
-\frac{1}{3}f_{\pi}\frac{m_{\pi}^{2}}{m_{u}+m_{d}}\phi_{3;\pi}^{(\sigma)}(\alpha)\leftrightarrow f_{\rho}^{T}m_{\rho}h_{||}^{(t)}(\alpha),\label{eq:tw3a_sub}
\end{equation}
\begin{equation}
f_{\pi}\frac{m_{\pi}^{2}}{m_{u}+m_{d}}\phi_{3;\pi}^{(p)}(\alpha)\leftrightarrow\left(f_{\rho}^{T}-f_{\rho}\frac{m_{u}+m_{d}}{m_{\rho}}\right)m_{\rho}h_{||}^{(s)}(\alpha).\label{eq:tw3b_sub}
\end{equation}

In Bjorken kinematics we expect that the dominant contribution
stems from the twist-two distributions $\phi_{2;\pi}$, ~$\phi_{2,\rho}^{(||)}$,
which might be decomposed as 
\begin{equation}
\phi_{2}\left(z,\,\,\mu^{2}\right)=6\,z\,\left(1-z\right)\left(1+\sum_{n>0}a_{2n}\left(\mu^{2}\right)C_{2n}^{3/2}\left(2z-1\right)\right),\label{eq:conformalExpansion}
\end{equation}
where the coefficients $a_{2n}\left(\mu^{2}\right)$ have mild multiplicative
dependence on the factorization scale $\mu$. 
The coefficients $a_{2n}$ are expected to be small, with current estimates~\cite{Fu:2016yzx,Bali:2017gfr}
\begin{align}
\left|a_{2}\left(\mu^{2}\approx2\,{\rm GeV}^{2}\right)\right|\sim\left|a_{4}\left(\mu^{2}\approx2\,{\rm GeV}^{2}\right)\right| & \lesssim0.1,\label{eq:a2n_Estimates}\\
\left|a_{2n}\left(\mu^{2}\approx2\,{\rm GeV}^{2}\right)\right|\approx0\quad{\rm for}\,\,\,n\geq3.
\end{align}
For this reason the ratio $R\left(x_{B},\,Q^{2}\right)$
defined in (\ref{eq:RRatio}) can be decomposed as
\begin{equation}
R\left(x_{B},\,Q^{2}\right)\approx\frac{f_{\rho}^{2}}{f_{\pi}^{2}}\left[1+2\,\sum_{n>0}r_{2n}\,\left(a_{2n,\rho}^{(||)}-a_{2n,\pi}\right)+\mathcal{O}\left(\left(a_{2,\rho}^{(||)}-a_{2,\pi}\right)^{2}\right)\right],\label{eq:R_exp}
\end{equation}
where the coefficients $r_{2n}$ correspond to the ratio of the DVMP
amplitudes evaluated with $n\not=0$ DAs, to the same amplitude evaluated
with $n=0$ (asymptotic) meson DAs. These coefficients will be analyzed
in Section~\ref{sec:Results}, considering their dependence on the implemented model
of GPDs. At next-to-leading order the coefficients $r_{2n}$ acquire
dependence on $x_{B}$, as well as a mild (logarithmic) dependence
on $Q^{2}$. The corrections to~(\ref{eq:R_exp}), due to higher twist
corrections, have a similar structure, although they decrease rapidly as
functions of virtuality, $\sim1/Q$. 

The twist-three distribution amplitudes of mesons contribute in the
combination $\phi_{3;p}\left(z,\,l_{\perp}\right)+2\phi_{3;\sigma}\left(z,\,l_{\perp}\right)$
(see Section~\ref{sec:DVMP_Xsec} for more details). For estimates
of the twist-3 contribution introduced in Section~\ref{sec:DVMP_Xsec},
we will use the parameterization suggested in~\cite{Goloskokov:2009ia,Goloskokov:2011rd},
\begin{equation}
\phi_{3}\left(z,\,l_{\perp}\right)=\phi_{3;p}\left(z,\,l_{\perp}\right)+2\phi_{3;\sigma}\left(z,\,l_{\perp}\right)=\frac{2a_{p}^{3}}{\pi^{3/2}}l_{\perp}\phi_{as}(z)\exp\left(-a_{p}^{2}l_{\perp}^{2}\right),\label{eq:phi_3}
\end{equation}
where the numerical constant $a_{p}$ is taken as $a_{p}\approx2\,{\rm GeV}^{-1}\approx0.4\,{\rm fm}$. 

\subsection{Leading twist evaluation}

\label{sec:DVMP_Xsec}
\begin{figure}[htp]
\includegraphics[height=3cm]{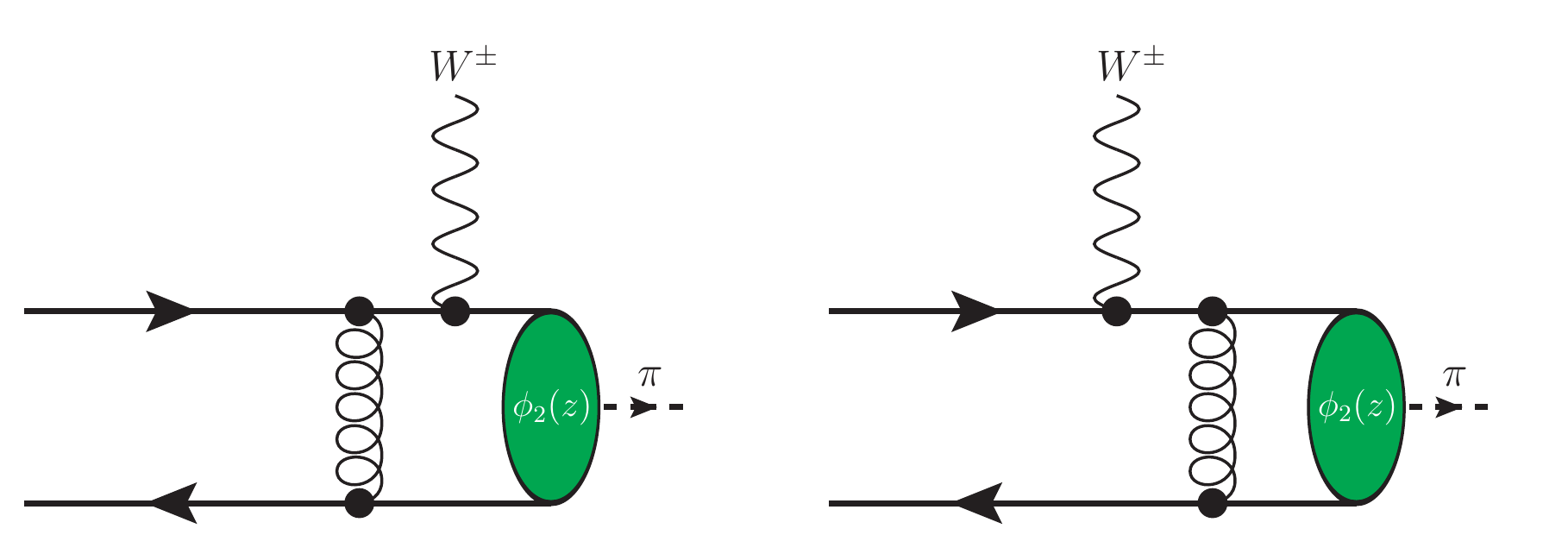} \includegraphics[height=3cm]{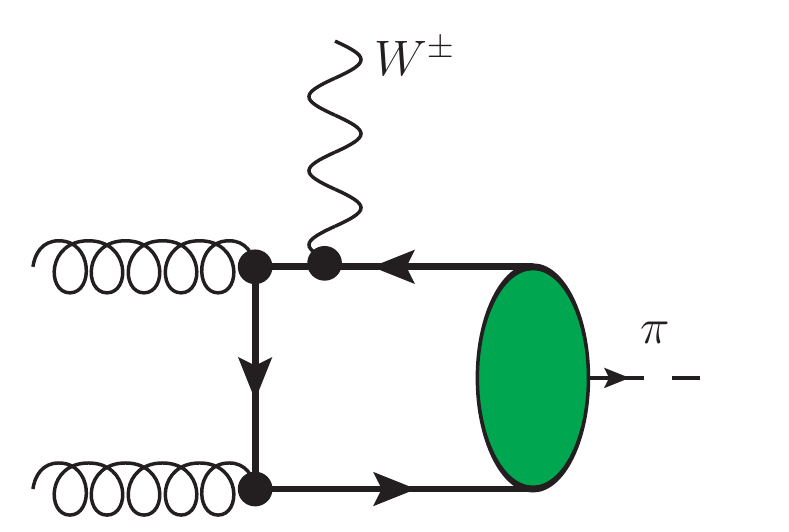}\protect\caption{\label{fig:DVMPLO}Leading-order contributions to the DVMP hard coefficient
functions. The green blob stands for the pion wave function. Additional
diagrams (not shown) may be obtained reversing directions of the quark
lines and in case of the last diagram, also permuting vector boson
vertices.}
\end{figure}

The CCDVMP might be studied both in neutrino-induced
and electron-induced processes. For the sake of definiteness, in what
follows we will consider the case of electroproduction, $ep\to\nu_{e}M\,p$.
The cross-section of this process is given by 
\begin{align}
\frac{d\sigma}{dt\,dx_{B}dQ^{2}} & =\Gamma\sum_{\nu\nu'}\mathcal{A}_{\nu',\nu L}^{*}\mathcal{A}_{\nu',\nu L},\label{eq:sigma_def}
\end{align}
where $t=\left(p_{2}-p_{1}\right)^{2}$ is the momentum transfer to
the proton, $Q^{2}=-q^{2}$ is the virtuality of the charged boson,
$x_{B}=Q^{2}/(2p\cdot q)$ is the Bjorken variable, the subscript
indices $\nu$ and $\nu'$ in the amplitude $\mathcal{A}$ refer to
helicity states of the baryon before and after interaction, and the
letter $L$ reflects the fact that in the Bjorken limit the dominant
contribution comes from the longitudinally polarized massive bosons
$W^{\pm}$~\cite{Ji:1998xh,Collins:1998be}. The kinematic factor
$\Gamma$ in~(\ref{eq:sigma_def}) for the charged current is given
explicitly by 
\begin{align}
\Gamma & =\frac{G_{F}^{2}\,x_{B}^{2}\left(1-y-\frac{\gamma^{2}y^{2}}{4}\right)}{64\pi^{3}Q^{2}\left(1+Q^{2}/M_{W}^{2}\right)^{2}\left(1+\gamma^{2}\right)^{3/2}},\label{eq:Pref}
\end{align}
where $\theta_{W}$ is the Weinberg angle, $M_{W}$ is the mass of
the heavy bosons $W^{\pm}$, $G_{F}$ is the Fermi constant, $f_{M}$
is the meson decay constant, and we also used the shorthand notations
\begin{equation}
\gamma=\frac{2\,m_{N}x_{B}}{Q},\quad y=\frac{Q^{2}}{s_{ep}\,x_{B}}=\frac{Q^{2}}{2m_{N}E_{e}\,x_{B}},\label{eq:elasticity}
\end{equation}
where $E_{e}$ is the electron energy in the target rest frame. In
Bjorken kinematics, the amplitude~$\mathcal{A}_{\nu',\nu L}$ factorizes
into a convolution of hard and soft parts, 
\begin{equation}
\mathcal{A}_{\nu',\nu}=\int_{-1}^{+1}dx\sum_{q=u,d,s,g}\,\sum_{\lambda\lambda'}\mathcal{H}_{\nu'\lambda',\nu\lambda}^{q}\mathcal{C}_{\lambda\lambda'}^{q},\label{eq:M_conv}
\end{equation}
where $x$ is the average light-cone fraction of the parton, superscript
$q$ is its flavor, $\lambda$ and $\lambda'$ are the helicities
of the initial and final partons, and $\mathcal{C}_{\lambda'\nu',\lambda\nu}^{q}$
is the hard coefficient function, which depends on the quantum numbers
of the produced meson and will be specified later. The soft matrix element
$\mathcal{H}_{\nu'\lambda',\nu\lambda}^{q}$ in~(\ref{eq:M_conv})
is diagonal in quark helicities ($\lambda,\,\lambda'$), and for the
twist-2 GPDs has a form 
\begin{align}
\mathcal{H}_{\nu'\lambda',\nu\lambda}^{q} & =\frac{2\delta_{\lambda\lambda'}}{\sqrt{1-\xi^{2}}}\left(-g_{A}^{q}\left(\begin{array}{cc}
\left(1-\xi^{2}\right)H^{q}-\xi^{2}E^{q} & \frac{\left(\Delta_{1}+i\Delta_{2}\right)E^{q}}{2m}\\
-\frac{\left(\Delta_{1}-i\Delta_{2}\right)E^{q}}{2m} & \left(1-\xi^{2}\right)H^{q}-\xi^{2}E^{q}
\end{array}\right)_{\nu'\nu}\right.\label{eq:HAmp}\\
 & +\left.{\rm sgn}(\lambda)g_{V}^{q}\left(\begin{array}{cc}
-\left(1-\xi^{2}\right)\tilde{H}^{q}+\xi^{2}\tilde{E}^{q} & \frac{\left(\Delta_{1}+i\Delta_{2}\right)\xi\tilde{E}^{q}}{2m}\\
\frac{\left(\Delta_{1}-i\Delta_{2}\right)\xi\tilde{E}^{q}}{2m} & \left(1-\xi^{2}\right)\tilde{H}^{q}-\xi^{2}\tilde{E}^{q}
\end{array}\right)_{\nu'\nu}\right),\nonumber 
\end{align}
where the constants $g_{V}^{q},\,g_{A}^{q}$ are the vector and axial
current couplings to quarks; the leading twist GPDs $H^{q},\,E^{q},\,\tilde{H}^{q}$
and $\tilde{E}^{q}$ are functions of variables $\left(x,\,\xi,\,t,\,\mu_{F}^{2}\right)$;
the skewness $\xi$ is related to the light-cone momenta of protons $p_{1,2}$
as $\xi=\left(p_{1}^{+}-p_{2}^{+}\right)/\left(p_{1}^{+}+p_{2}^{+}\right)$;
the invariant momentum transfer $t=\Delta^{2}=\left(p_{2}-p_{1}\right)^{2}$,
and $\mu_{F}$ is the factorization scale (see e.g.~\cite{Goeke:2001tz,Diehl:2003ny}
for details of the kinematics).  The evaluation of the structure
function $\mathcal{C}^{q}$ is quite straightforward, and in leading
order over $\alpha_{s}$ it gets contributions from the diagrams shown
schematically in Figure~\ref{fig:DVMPLO}. This has been studied both
for pion electroproduction~\cite{Vanderhaeghen:1998uc,Mankiewicz:1998kg,Goloskokov:2006hr,Goloskokov:2007nt,Goloskokov:2008ib,Goloskokov:2011rd,Goldstein:2012az}
and neutrinoproduction~\cite{Kopeliovich:2012dr}. For the processes
in which baryon does not change its internal state, there are additional
contributions from gluon GPDs, as shown in the rightmost panel of
the Figure~\ref{fig:DVMPLO}. These corrections are small in JLAB
kinematics, yet give a sizable contribution at higher energies. In
the next-to-leading order, the coefficient function includes an additional
gluon attached in all possible ways to all diagrams in Figure~\ref{fig:DVMPLO},
as well as additional contributions from sea quarks, as shown in the
Figure~\ref{fig:DVMPNLO-1}.

\begin{quote}
\begin{figure}[htp]
\includegraphics[height=3cm]{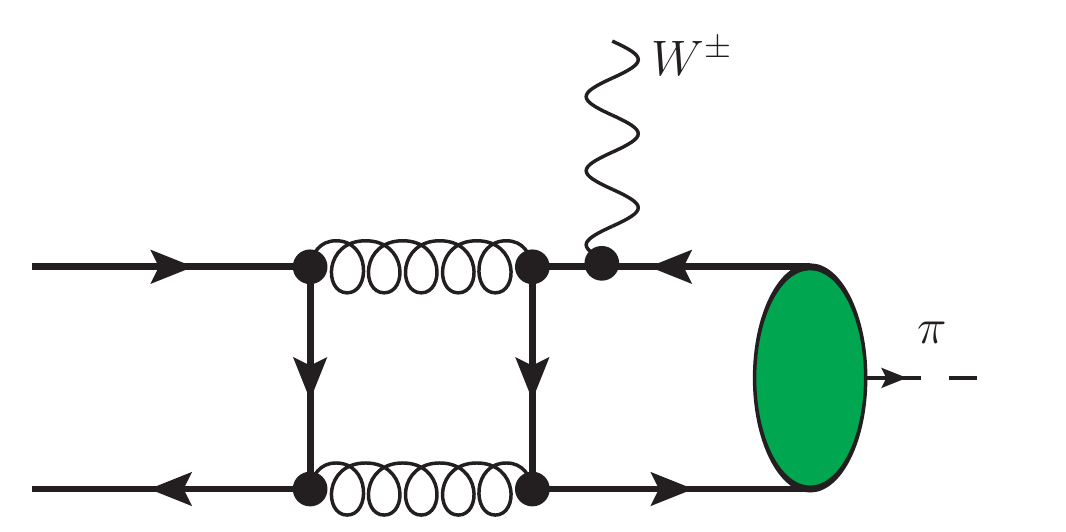}\protect\caption{\label{fig:DVMPNLO-1}Sea quark contributions to the DVMP, which appear
at next-to-leading-order. Additional diagrams (not
shown) may be obtained reversing directions of the quark lines.}
\end{figure}
Straightforward evaluation of the diagrams shown in the Figures~\ref{fig:DVMPLO},\ref{fig:DVMPNLO-1}
yields for the coefficient function
\begin{align}
\mathcal{C}_{\lambda\lambda'}^{q} & =\delta_{\lambda\lambda'}\left(\eta_{-}^{q}c_{-}^{(q)}\left(x,\,\xi\right)+{\rm sgn}(\lambda)\eta_{+}^{q}c_{+}^{(q)}\left(x,\,\xi\right)+\mathcal{O}\left(\frac{m^{2}}{Q^{2}}\right)+\mathcal{O}\left(\alpha_{s}^{2}\left(\mu_{R}^{2}\right)\right)\right),\label{eq:Coef_function}
\end{align}
where the process-dependent flavor factors $\eta_{V\pm}^{q},\,\eta_{A\pm}^{q}$
are the same for $J^{P}=0^{-}$- and $1^{-}$mesons, and are given
explicitly in Table~\ref{tab:DVMP_amps}~\footnote{As was discussed above, for processes with change of internal baryon
structure, we use $SU(3)$ relations~\cite{Frankfurt:1999fp}, which
are valid up to corrections in current quark masses $\sim\mathcal{O}\left(m_{q}\right)$.}. Also, in~(\ref{eq:Coef_function}) we introduced the shorthand notation
\end{quote}
\begin{table}
\protect\caption{\label{tab:DVMP_amps}The flavor coefficients $\eta_{\pm}^{q}$ for
several meson production processes discussed in this paper. We use the
notation $q=u,d,s,...$ $\{M^{\pm,0},M_{s}^{\pm,0}\}=\{\pi^{\pm,0},\,K^{\pm,0}\}$
mesons in $J^{P}=0^{-}$multiplet, and $\{M^{\pm,0},M_{s}^{\pm,0}\}=\{\rho^{\pm,0},\,K^{*\pm,0}\}$
mesons in $J^{P}=1^{-}$multiplet. As commented in the text, CC currents
could be studied either in electron-induced processes (so $\{\ell,\ell'\}=\{e,\,\nu_{e}\}$)
or in neutrino-induced processes, $\{\ell,\,\ell'\}=\{\bar{\nu}_{e},\,e^{+}\}$.
For the case of CC mediated processes, the $V-A$ structure of the
charged current implies $\eta_{V\pm}^{q}=\eta_{\pm}^{q},\quad\eta_{A\pm}^{q}=-\eta_{\pm}^{q}$.}

\global\long\def\arraystretch{1.5}

\begin{tabular}{|c|c|c|c|c|c|c|}
\cline{1-3} \cline{5-7} 
Process  & $\eta_{+}^{q}$  & $\eta_{-}^{q}$  &  & Process  & $\eta_{+}^{q}$  & $\eta_{-}^{q}$\tabularnewline
\cline{1-3} \cline{5-7} 
$\ell\,p\to\ell'\,M^{-}p$  & $V_{ud}\delta_{qd}$  & $V_{ud}\delta_{qu}$  &  & $\ell\,p\to\ell'M^{0}n$  & $V_{ud}\frac{\delta_{qu}-\delta_{qd}}{\sqrt{2}}$  & $-V_{ud}\frac{\delta_{qu}-\delta_{qd}}{\sqrt{2}}$ \tabularnewline
\cline{1-3} \cline{5-7} 
$\ell\,p\to\ell'M^{0}n$  & $V_{ud}\frac{\delta_{qu}-\delta_{qd}}{\sqrt{2}}$  & $-V_{ud}\frac{\delta_{qu}-\delta_{qd}}{\sqrt{2}}$  &  & $\ell\,p\to\ell'M^{-}n$  & $V_{ud}\delta_{qu}$  & $V_{ud}\delta_{qd}$\tabularnewline
\cline{1-3} \cline{5-7} 
$\ell\,p\to\ell'M_{s}^{-}p$  & $V_{us}\delta_{qs}$  & $V_{us}\delta_{qs}$  &  & $\ell\,n\to\ell'M_{s}^{0}\Sigma^{-}$  & 0 & $-V_{ud}\left(\delta_{qu}-\delta_{qs}\right)$ \tabularnewline
\cline{1-3} \cline{5-7} 
\end{tabular}
\end{table}

\begin{eqnarray}
c_{\pm}^{(q)}\left(x,\xi\right) & = & \frac{8\pi i}{9}\frac{\alpha_{s}\left(\mu_{R}^{2}\right)f_{M}}{Q}\frac{1}{x\pm\xi\mp i0}\int_{0}^{1}dz\,\frac{\phi_{2}\left(z\right)}{z}\left(1+\frac{\alpha_{s}\left(\mu_{r}^{2}\right)}{2\pi}T^{(1)}\left(\frac{\xi\pm x}{2\xi},\,z\right)\right),\label{eq:c2}
\end{eqnarray}
where $\phi_{2}(z)$ is the twist-2 meson distribution amplitude (DA).
The function $T^{(1)}\left(v,\,z\right)$ in~(\ref{eq:c2}) encodes
NLO corrections to the coefficient function and is given explicitly
in the Appendix~\ref{sec:NLOCoef}. In general, we could expect that
the spin structure of the coefficient function $\mathcal{C}_{\lambda}^{q}$
should depend on the quantum numbers of the produced mesons, however in
the leading twist this is not so. This happens because at leading
twist the distribution amplitudes of the $J^{P}=0^{-}$ and $1^{-}$
mesons differ only by an additional $\gamma_{5}$ in the corresponding quark
operator and $V-A$ structure of charged current. From a trivial identity
\begin{equation}
\gamma_{\mu}\left(1-\gamma_{5}\right)\gamma_{\alpha_{1}}S\left(p_{1}\right)...\gamma_{\alpha_{n}}S\left(p_{n}\right)\gamma_{\pm}\gamma_{5}=\gamma_{\mu}\left(1-\gamma_{5}\right)\gamma_{\alpha_{1}}S\left(p_{1}\right)...\gamma_{\alpha_{n}}S\left(p_{n}\right)\gamma_{\pm}\label{eq:Identity}
\end{equation}
where $S\left(p_{i}\right)$ are the quark propagators (massless in the
Bjorken limit), we may conclude that for charged currents the amplitudes
of $\rho$- and $\pi$-production coincide \emph{to any order} in
the strong coupling constant $\alpha_{s}\left(Q^{2}\right)$\footnote{For neutral currents this statement is not valid due to differences
in vector and axial charges, $g_{V}\not=g_{A}$.}. The corrections due to finite mass of the quarks are $\sim\mathcal{O}(m_{q}/Q)$,
and are numerically negligible for light quarks. In the twist-three
case, similar arguments hold for the two-parton distribution amplitudes,
yet for the contributions of the three-parton DAs this is no longer
so. For this reason, we may use the above-mentioned substitutions~(\ref{eq:tw2_sub},\ref{eq:tw3a_sub},\ref{eq:tw3b_sub})
to relate the pion and $\rho$-meson distribution amplitudes.

In the leading order over $\alpha_{s}$, the ratio $R_{\rho/\pi}$
defined in~(\ref{eq:RRatio}) is constant and is given by the ratio
of the minus-first moments $\left\langle \phi_{2,\,||,\rho}^{-1}\right\rangle $
and $\left\langle \phi_{2,\pi}^{-1}\right\rangle $. In terms of the
conformal expansion coefficients $a_{2n}$ defined in~(\ref{eq:conformalExpansion}),
the moments may be evaluated exactly and are given by $\left\langle \phi_{2}^{-1}\right\rangle =1+\sum_{n}a_{2n}$,
so the ratio~(\ref{eq:RRatio}) is given by 
\begin{equation}
R_{\rho/\pi}\approx\left(f_{\rho}/f_{\pi}\right)^{2}\left(\frac{1+\sum a_{2n,\rho}^{(||)}}{1+\sum a_{2n,\pi}}\right)^{2}.
\end{equation}
At this order all the expansion coefficients $r_{2n}$ defined in~(\ref{eq:R_exp}) are
equal to unity, $r_{2n}\left(x_{B},\,Q^{2}\right)=1$, and do not
depend on $\left(x_{B},\,Q^{2}\right)$. In the next-to-leading order
there are $\delta r_{2n}\sim\mathcal{O}\left(\alpha_{s}\right)$ corrections,
given explicitly in Appendix~~\ref{sec:NLOCoef}. The numerical
values of the coefficients are discussed in detail in the following
Section~\ref{sec:Results}.

\subsection{Twist-three corrections}

\label{sec:Tw3}In the Bjorken limit, it is expected that the dominant
contribution should come from the twist-two GPDs $H,\,E,\,\tilde{H},\,\tilde{E}$.
However, as was shown in~~\cite{Defurne:2016eiy}, in moderate-energy
experiments the typical values\lyxdeleted{Linux User}{Mon Apr  1 12:54:54 2019}{
} of virtuality $Q$ are only two or three times larger than the\lyxdeleted{Linux User}{Mon Apr  1 12:54:54 2019}{
} mass of the nucleon $m_{N}$. For this reason it is important to
assess how large are the omitted higher-twist contributions. 

Technically the evaluation of the twist-three contributions is quite
challenging, because the are many different contributions, and for
some of them (see e.g. three-parton contributions analyzed in~\cite{Anikin:2009bf,Anikin:2009hk})
numerical estimates are currently challenging due to lack of reliable
phenomenological restrictions on multiparton distributions. In this
paper we will restrict ourselves to the estimates of higher twist
contributions due to two-parton twist-three components of the meson
wave functions, which are expected to give the largest contribution
to the difference between pion and $\rho$-meson cross-sections. The
corresponding twist-three DAs for pion and $\rho$-meson were defined
in Section~\ref{subsec:DAs}. Previously this analysis has been done
by us in the context of neutrino-production \cite{Kopeliovich:2014pea}
and pion production by charged currents~\cite{Siddikov:2017nku},
and here we briefly repeat it for the case of charged current meson
production. For the case of $\rho$-meson the amplitudes might be
obtained from pion amplitude by the substitution~(\ref{eq:tw3a_sub},~\ref{eq:tw3b_sub}).
The twist-three meson DAs probe the so-called transversity GPDs, which
contribute to the amplitude~(\ref{eq:HAmp}) as\lyxdeleted{Linux User}{Mon Apr  1 12:54:54 2019}{
} 
\begin{align}
\delta\mathcal{H}_{\nu'\lambda',\nu\lambda}^{q} & =\left(m_{\nu'\nu}^{q}\delta_{\lambda,-}\delta_{\lambda',+}+n_{\nu'\nu}^{q}\delta_{\lambda,+}\delta_{\lambda',-}\right),
\end{align}
where the coefficients $m_{\pm,\pm}^{q}$ and $n_{\pm,\pm}^{q}$ are
linear combinations of the transversity GPDs, 
\begin{align}
m_{--}^{q} & =\frac{\sqrt{-t'}}{4m}\left[2\tilde{H}_{T}^{q}\,+(1+\xi)E_{T}^{q}-(1+\xi)\tilde{E}_{T}^{q}\right],\\
m_{-+}^{q} & =\sqrt{1-\xi^{2}}\frac{t'}{4m^{2}}\tilde{H}_{T}^{q},\\
m_{+-}^{q} & =\sqrt{1-\xi^{2}}\left[H_{T}^{q}-\frac{\xi^{2}}{1-\xi^{2}}E_{T}^{q}+\frac{\xi}{1-\xi^{2}}\tilde{E}_{T}^{q}-\frac{t'}{4m^{2}}\tilde{H}_{T}^{q}\right],\\
m_{++}^{q} & =\frac{\sqrt{-t'}}{4m}\left[2\tilde{H}_{T}^{q}+(1-\xi)E_{T}^{q}+(1-\xi)\tilde{E}_{T}^{q}\right],
\end{align}

\begin{align}
n_{--}^{q} & =-\frac{\sqrt{-t'}}{4m}\left(2\tilde{H}_{T}^{q}+(1-\xi)E_{T}^{q}+(1-\xi)\tilde{E}_{T}^{q}\right),\\
n_{-+}^{q} & =\sqrt{1-\xi^{2}}\left(H_{T}^{q}-\frac{\xi^{2}}{1-\xi^{2}}E_{T}^{q}+\frac{\xi}{1-\xi^{2}}\tilde{E}_{T}^{q}-\frac{t'}{4m^{2}}\tilde{H}_{T}^{q}\right),\\
n_{+-}^{q} & =\sqrt{1-\xi^{2}}\frac{t'}{4m^{2}}\tilde{H}_{T}^{q},\\
n_{++}^{q} & =-\frac{\sqrt{-t'}}{4m}\left(2\tilde{H}_{T}^{q}+(1+\xi)E_{T}^{q}-(1+\xi)\tilde{E}_{T}^{q}\right),
\end{align}
and we introduced a shorthand notation $t'=-\Delta_{\perp}^{2}/(1-\xi^{2})$;
$\Delta_{\perp}=p_{2,\perp}-p_{1,\perp}$ is the transverse part of
the momentum transfer. The coefficient function~(\ref{eq:Coef_function})
also gets an additional nondiagonal in parton helicity contribution,

\begin{align}
\delta\mathcal{C}_{\lambda'0,\lambda\mu}^{q} & ==\delta_{\mu,+}\delta_{\lambda,-}\delta_{\lambda',+}\left(S_{A}^{q}-S_{V}^{q}\right)+\delta_{\mu,-}\delta_{\lambda,+}\delta_{\lambda',-}\left(S_{A}^{q}+S_{V}^{q}\right)+\mathcal{O}\left(\frac{m^{2}}{Q^{2}}\right),\label{eq:Coef_function-1}
\end{align}
where we introduced the shorthand notations
\begin{eqnarray}
S_{A}^{q} & = & \int dz\,\left(\left(\eta_{A+}^{q}c_{+}^{(3,p)}\left(x,\xi\right)-\eta_{A-}^{q}c_{-}^{(3,p)}\left(x,\xi\right)\right)+2\left(\eta_{A-}^{q}c_{-}^{(3,\sigma)}\left(x,\xi\right)+\eta_{A+}^{q}c_{+}^{(3,\sigma)}\left(x,\xi\right)\right)\right),\label{eq:SA_def}\\
S_{V}^{q} & = & \int dz\,\left(\left(\eta_{V+}^{q}c_{+}^{(3,p)}\left(x,\xi\right)+\eta_{V-}^{q}c_{-}^{(3,p)}\left(x,\xi\right)\right)+2\left(\eta_{V+}^{q}c_{+}^{(3,\sigma)}\left(x,\xi\right)-\eta_{V-}^{q}c_{-}^{(3,\sigma)}\left(x,\xi\right)\right)\right),\label{eq:SV_def}
\end{eqnarray}
\begin{equation}
c_{+}^{(3,i)}\left(x,\xi\right)=\frac{4\pi i\alpha_{s}f_{\pi}\xi}{9\,Q^{2}}\int_{0}^{1}dz\frac{\phi_{3,i}(z)}{z\,(x+\xi)^{2}},\quad c_{-}^{(3,i)}\left(x,\xi\right)=\frac{4\pi i\alpha_{s}f_{\pi}\xi}{9\,Q^{2}}\int_{0}^{1}dz\frac{\phi_{3,i}(z)}{(1-z)(x-\xi)^{2}};\label{eq:Tw3_coefFunction}
\end{equation}
and the twist-three pion distributions are defined in Section~\ref{subsec:DAs}.
Due to symmetry of $\phi_{p}$ and antisymmetry of $\phi_{\sigma}$
with respect to charge conjugation, the dependence on the pion DAs
factorizes in the collinear approximation and contributes only as
the minus first moment of the linear combination of the twist-3 DAs,
$\phi_{3}^{(p)}(z)+2\phi_{3}^{(\sigma)}(z)$, 
\begin{equation}
\left\langle \phi_{3}^{-1}\right\rangle =\int_{0}^{1}dz\frac{\phi_{3}^{(p)}\left(z\right)+2\phi_{3}^{(\sigma)}\left(z\right)}{z}.
\end{equation}
In general case the coefficient function ~(\ref{eq:Tw3_coefFunction})
leads to collinear divergencies near the points $x=\pm\xi$, when substituted
to (\ref{eq:M_conv}). As was noted in~\cite{Goloskokov:2009ia},
this singularity is naturally regularized by the small transverse
momentum of the quarks inside the meson. Such regularization modifies~(\ref{eq:Tw3_coefFunction})
to 
\begin{align}
c_{+}^{(3,i)}\left(x,\xi\right) & =\frac{4\pi i\alpha_{s}f_{\pi}\xi}{9\,Q^{2}}\int_{0}^{1}dz\,d^{2}l_{\perp}\frac{\phi_{3,i}\left(z,\,l_{\perp}\right)}{(x+\xi-i0)\left(z(x+\xi)+\frac{2\xi\,l_{\perp}^{2}}{Q^{2}}\right)},\label{eq:c3Plus}\\
c_{-}^{(3,i)}\left(x,\xi\right) & =\frac{4\pi i\alpha_{s}f_{\pi}\xi}{9\,Q^{2}}\int_{0}^{1}dz\,d^{2}l_{\perp}\frac{\phi_{3,i}\left(z,\,l_{\perp}\right)}{(x-\xi+i0)\left((1-z)(x-\xi)-\frac{2\xi\,l_{\perp}^{2}}{Q^{2}}\right)},\label{eq:c3Minus}
\end{align}
where $l_{\perp}$ is the transverse momentum of the quark, and we
tacitly assume absence of any other transverse momenta in the coefficient
function. Due to interference of the leading twist and twist-three
contributions, the total cross-section acquires dependence on the
angle $\varphi$ between lepton scattering and pion production planes,
\begin{align}
\frac{d\sigma}{dt\,dx_{B}dQ^{2}d\varphi} & =\epsilon\frac{d\sigma_{L}}{dt\,dx_{B}dQ^{2}d\varphi}+\frac{d\sigma_{T}}{dt\,dx_{B}dQ^{2}d\varphi}+\sqrt{\epsilon(1+\epsilon)}\cos\varphi\frac{d\sigma_{LT}}{dt\,dx_{B}dQ^{2}d\varphi}\label{eq:sigma_def-1}\\
 & +\epsilon\cos\left(2\varphi\right)\frac{d\sigma_{TT}}{dt\,dx_{B}dQ^{2}d\varphi}+\sqrt{\epsilon(1+\epsilon)}\sin\varphi\frac{d\sigma_{L'T}}{dt\,dx_{B}dQ^{2}d\varphi}+\epsilon\sin\left(2\varphi\right)\frac{d\sigma_{T'T}}{dt\,dx_{B}dQ^{2}d\varphi},\nonumber 
\end{align}
where we introduced the shorthand notations
\begin{equation}
\epsilon=\frac{1-y-\frac{\gamma^{2}y^{2}}{4}}{1-y+\frac{y^{2}}{2}+\frac{\gamma^{2}y^{2}}{4}}.
\end{equation}
\begin{align}
\frac{d\sigma_{L}}{dt\,dx_{B}dQ^{2}d\varphi} & =\frac{\Gamma\,\sigma_{00}}{2\pi\epsilon}\label{eq:sigma_L}\\
\frac{d\sigma_{T}}{dt\,dx_{B}dQ^{2}d\varphi} & =\frac{\Gamma}{2\pi\epsilon}\,\left(\frac{\sigma_{++}+\sigma_{--}}{2}+\frac{1}{2}\sqrt{1-\epsilon^{2}}\frac{\sigma_{++}-\sigma_{--}}{2}\right)\label{eq:sigma_T}\\
\frac{d\sigma_{LT}}{dt\,dx_{B}dQ^{2}d\varphi} & =\frac{\Gamma}{2\pi\epsilon}\,\left({\rm Re}\left(\sigma_{0+}-\sigma_{0-}\right)+\frac{1}{2}\sqrt{\frac{1-\epsilon}{1+\epsilon}}{\rm Re}\left(\sigma_{0+}+\sigma_{0-}\right)\right)\label{eq:sigma_LT}\\
\frac{d\sigma_{TT}}{dt\,dx_{B}dQ^{2}d\varphi} & =-\frac{\Gamma}{2\pi\epsilon}\,{\rm Re}\left(\sigma_{+-}\right)\label{eq:sigma_TT}\\
\frac{d\sigma_{L'T}}{dt\,dx_{B}dQ^{2}d\varphi} & =-\frac{\Gamma}{2\pi\epsilon}\,\left({\rm Im}\left(\sigma_{+0}+\sigma_{-0}\right)-\frac{1}{2}\sqrt{\frac{1-\epsilon}{1+\epsilon}}{\rm Im}\left(\sigma_{-0}-\sigma_{+0}\right)\right)\label{eq:sigma_LPrimeT}\\
\frac{d\sigma_{T'T}}{dt\,dx_{B}dQ^{2}d\varphi} & =-\frac{\Gamma}{2\pi\epsilon}\,{\rm Im}\left(\sigma_{+-}\right)\label{eq:sigma_TPrimeT}
\end{align}
and the subindices $\alpha,\beta$ in 
\begin{equation}
\sigma_{\alpha\beta}=\sum_{\nu\nu'}\mathcal{A}_{\nu'0,\nu\alpha}^{*}\mathcal{A}_{\nu'0,\nu\beta},
\end{equation}
refer to the polarizations of intermediate heavy boson in the amplitude
and its conjugate. As we will see below, in JLAB kinematics the contribution
of higher twist corrections is small, and for this reason we will quantify
their size in terms of the angular harmonics $c_{n},\,s_{n}$, normalizing
the total cross-section to the cross-section of the dominant DVMP
process defined as~\cite{Siddikov:2017nku}
\begin{equation}
\frac{d^{4}\sigma^{(tot)}}{dt\,d\ln x_{Bj}\,dQ^{2}d\varphi}=\frac{1}{2\pi}\frac{d^{4}\sigma^{(DVMP)}}{dt\,d\ln x_{Bj}\,dQ^{2}}\left(1+\sum_{n=0}^{2}c_{n}\cos(n\varphi)+s_{1}\sin(\varphi)\right).\label{eq:Harmonics}
\end{equation}
The main purpose of this study is to analyze the sensitivity of the
ratio~(\ref{eq:RRatio}) to changes of the coefficients $r_{2n}$.
For this reason in what follows we will focus on the evaluation of the
harmonics $c_{0}$ and the corresponding cross-sections $d\sigma_{L}$
and $d\sigma_{T}$. The higher twist corrections contribute additively
to the cross-section (no interference due to different spin structure),
and as we will see below, in the kinematics of interest the cross-section
$d\sigma_{T}\ll d\sigma_{L}$. For this reason the correction to the
ratio~(\ref{eq:RRatio}) is small and is given by
\begin{align}
\delta R^{{\rm twist}-3} & \approx\underbrace{\frac{f_{\rho}^{2}}{f_{\pi}^{2}}\left(\frac{d\sigma_{T}^{(\rho)}}{\epsilon d\sigma_{L}^{(\rho)}}-\frac{d\sigma_{T}^{(\pi)}}{\epsilon d\sigma_{L}^{(\pi)}}\right)}_{\sim\mathcal{O}(1/Q)}+\mathcal{O}\left(\frac{1}{Q^{2}}\right)=\frac{f_{\rho}^{2}}{f_{\pi}^{2}}\left(c_{0,\rho}-c_{0,\pi}\right)+\mathcal{O}\left(\frac{1}{Q^{2}}\right):=\frac{f_{\rho}^{2}}{f_{\pi}^{2}}\Delta c_{0},\label{eq:dR3}\\
\Delta c_{0} & =c_{0,\rho}-c_{0,\pi},
\end{align}
where $c_{0,\rho}$ and $c_{0,\pi}$ are the zeroth order harmonics
(angular-independent contributions of twist-3 terms) of the $\rho$-meson
and pion respectively. At present, the values of the twist-three $\rho$-meson
DAs are poorly known (especially for the case of $\rho$-mesons), and
for this reason we will assume that it changes from 0 up to the same
value as for pion, (\ref{eq:phi_3}).

\section{Results and discussion}

\label{sec:Results}

In this section we would like to present numerical results for the
charged current pion production. For the sake of definiteness, for
numerical estimates we use the Kroll-Goloskokov parameterization of
GPDs~\cite{Goloskokov:2006hr,Goloskokov:2007nt,Goloskokov:2008ib,Goloskokov:2009ia,Goloskokov:2011rd}.
For illustration, we will start the discussion assuming
dominance of the twist two corrections, and neglecting the deviations
from the asymptotic form encoded in the coefficients $a_{2n}$ in~(\ref{eq:conformalExpansion}).
In this case the difference between pion and $\rho$-meson cross-sections
becomes negligible (we may neglect the so-called ``kinematic'' higher
twist effects $\sim\mathcal{O}\left(M_{\pi,\rho}^{2}/Q^{2}\right)$
in the Bjorken limit). 

\begin{figure}
\includegraphics[height=5cm]{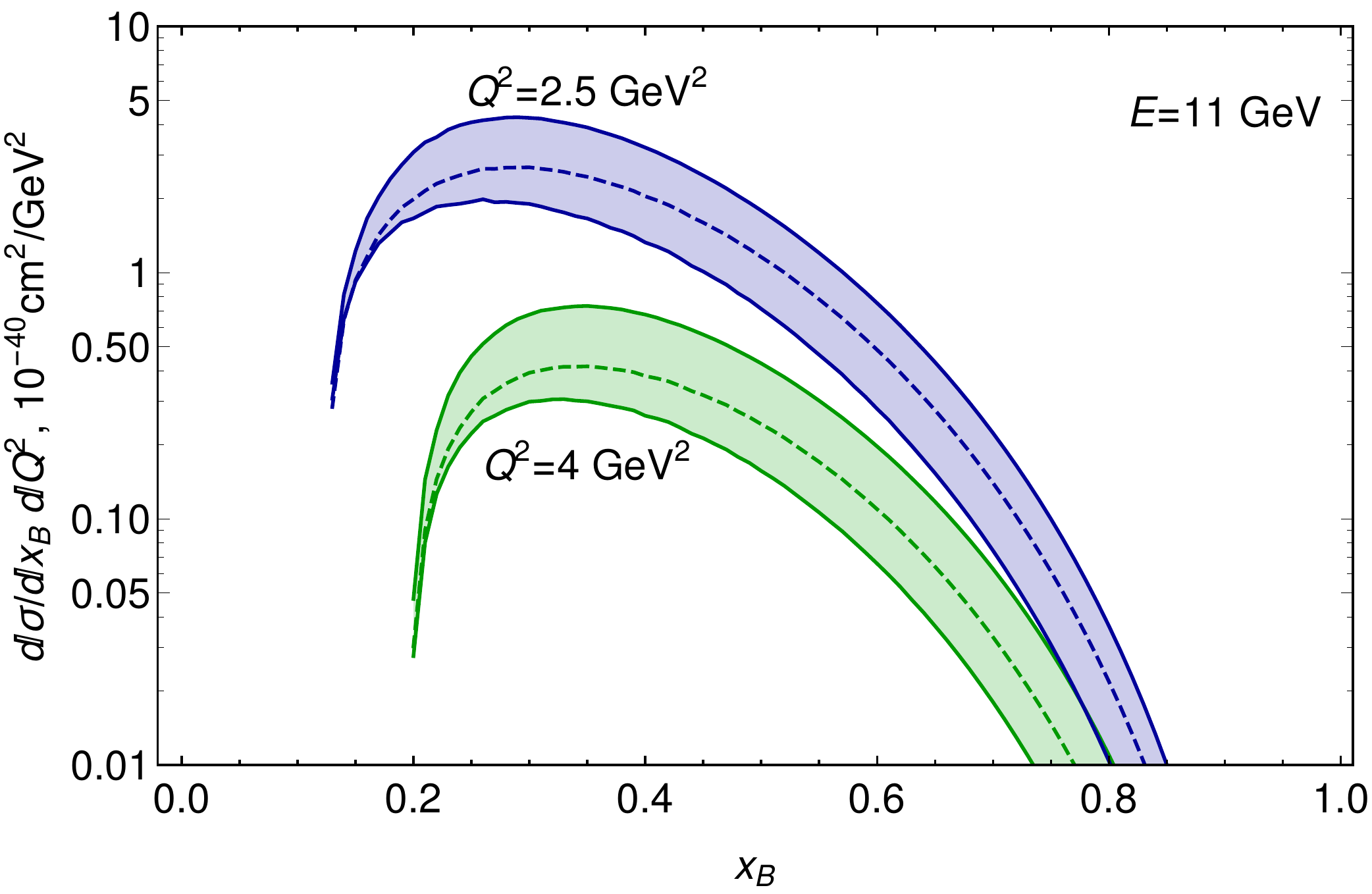}\includegraphics[height=5cm]{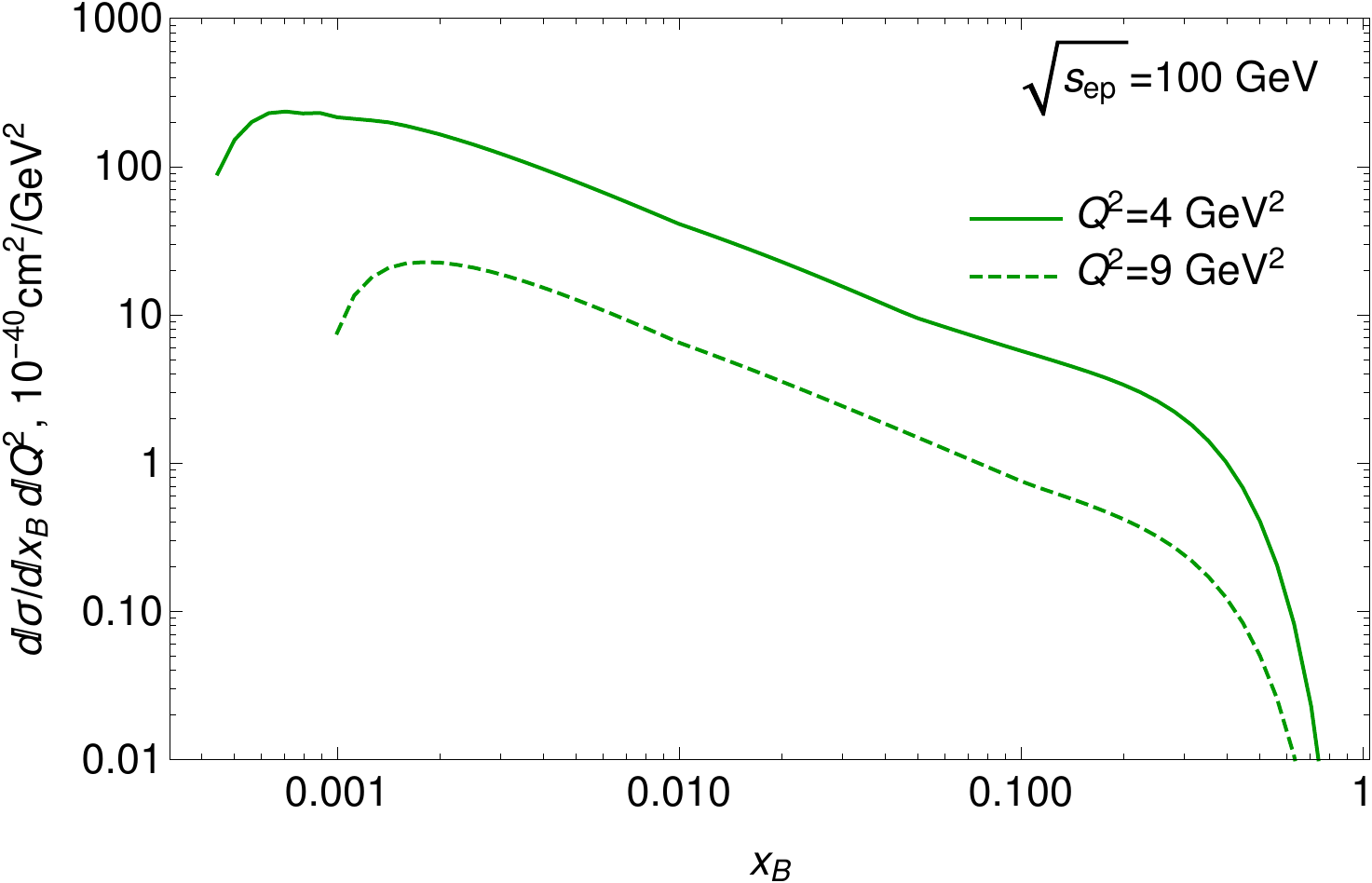}

\protect\caption{\label{fig:DVMP-pions}(color online) Left plot: Charged current meson
production cross-section on a proton target, within JLab kinematics (fixed
electron energy $E=11\,{\rm GeV}$). Evaluations are performed using
NLO coefficient functions, as discussed in Section~\ref{sec:DVMP_Xsec}.
The width of the band represents the uncertainty due to the factorization
scale choice~$\mu_{F}\in\left(Q/2,\,2Q\right)$, as explained in
the text. Right plot: $x_{B}$-dependence of the cross-section in
EIC kinematics with $\sqrt{s}_{ep}\approx100\,{\rm GeV}$. For other
values of $\sqrt{s}_{ep}$ and fixed $\left(x_{B},\,Q^{2}\right)$
the cross-section might be obtained by rescaling the factor~(\ref{eq:Pref}).
This factor is responsible for the suppression of the cross-section at
small $x_{B}\ll1$ and at fixed energy $\sqrt{s_{ep}}$.}
\end{figure}

In the left panel of the Figure~\ref{fig:DVMP-pions} we show
predictions for the differential cross-section $d\sigma/dx_{B}\,dQ^{2}$
for charged meson ($\rho^{-},\,\pi^{-}$) production, within JLab kinematics.
We expect that for typical instant luminosities $\sim10^{35}{\rm cm^{-2}s^{-1}}$,
easonable statistics could be collected after 30-60 days of running.
At fixed electron energy $E_{e}=11\,{\rm GeV}$ and virtuality $Q^{2}$,
the cross-section as function of $x_{B}$ has a typical bump-like
shape, which is explained by an interplay of two factors. For small
$x_{B}\sim Q^{2}/2m_{N}E_{e}$ the elasticity $y$ defined in~(\ref{eq:elasticity})
approaches one, which causes a suppression due to a prefactor $\Gamma$
in~(\ref{eq:sigma_def}). In the opposite limit, the suppression
$\sim(1-x)^{n}$ is due to the implemented parameterization of GPDs.
In the evaluation of the coefficient function we take into account NLO
corrections, which give a sizable contribution for $Q^{2}\lesssim10\,{\rm GeV}^{2}$.
The band around the curves reflects the uncertainty of the predictions
due to higher order corrections, which was obtained varying the factorization
scale $\mu_{F}$ in the range~$\mu_{F}\in\left(Q/2,\,2Q\right)$
(see~\cite{Diehl:2003ny,Goloskokov:2009ia,Goloskokov:2011rd,Diehl:2007hd,Pire:2017lfj}
for more details). The amplitudes in this region get the dominant
contribution from the GPDs $H^{u},\,H^{d}$, whereas helicity flip
and gluon GPDs give a minor ($\sim$10\%) correction to the full cross-section.
In the right panel we show the cross-section for the kinematics
of EIC experiment, assuming a center-of-mass energy $\sqrt{s_{ep}}\approx100\,{\rm GeV}.$
At present the exact energy $\sqrt{s_{ep}}$, which will be available
at EIC, is not known, yet reevaluation for other energies $\sqrt{s}_{ep}$
is quite straightforward and might be obtained by rescaling the $y$-dependent
prefactor~(\ref{eq:Pref}). The effects of this factor are pronounced
at small $x_{B}\ll1$, where it leads to a suppression of the cross-section.

\begin{figure}
\includegraphics[height=6cm]{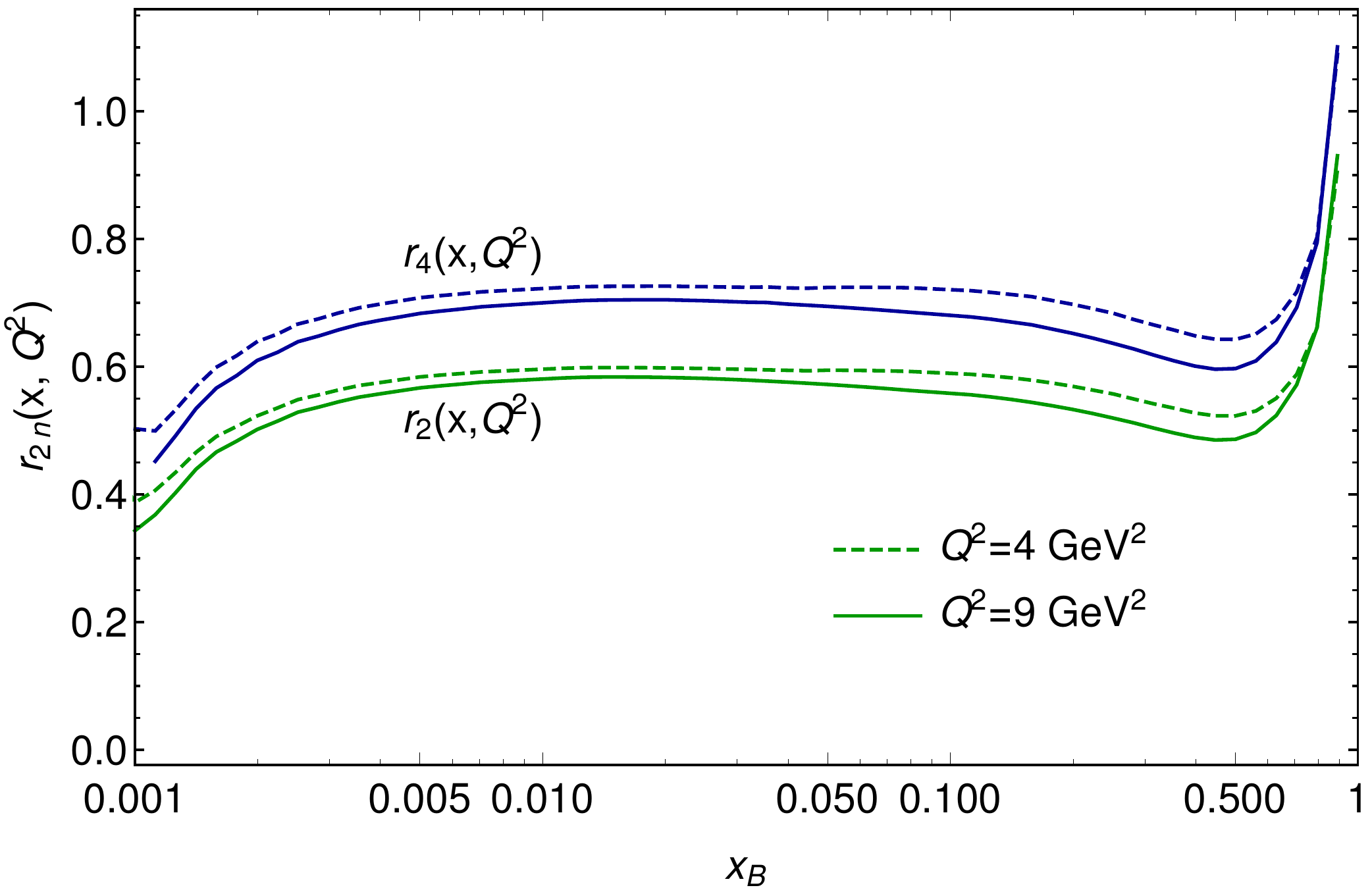}\includegraphics[height=6cm]{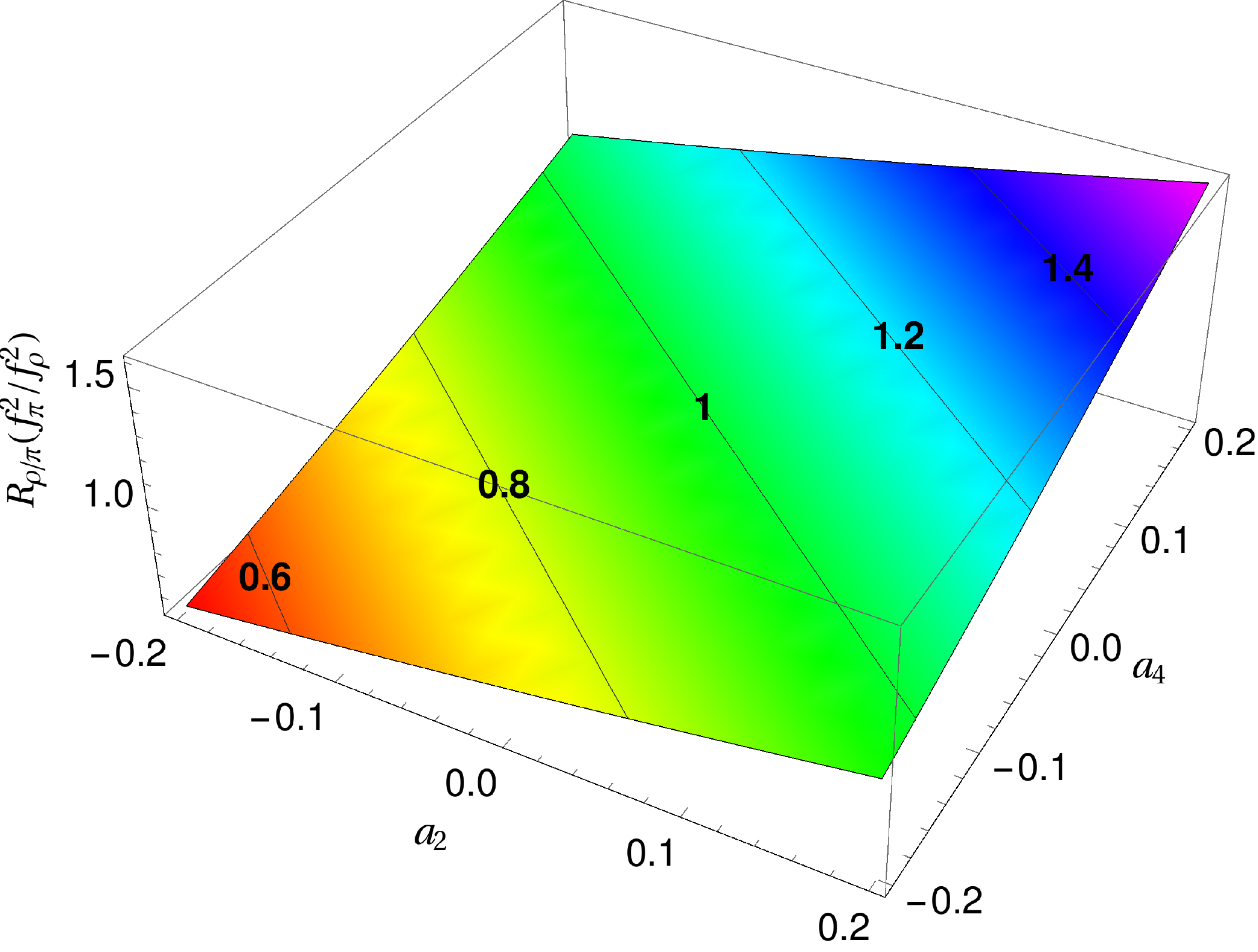}

\protect\caption{\label{fig:DVMP-r2n}(color online) Left: Values of the coefficients
$r_{2n}\left(x_{B},\,Q^{2}\right).$ The two bottom curves correspond
to $r_{2}\left(x_{B},\,Q^{2}\right);$ the two upper curves correspond
to $r_{4}\left(x_{B},\,Q^{2}\right).$ For both cases dashed lines
correspond to $Q^{2}=4\,{\rm GeV}^{2}$, solid lines correspond to
$Q^{2}=9\,{\rm GeV}^{2}.$ All evaluations performed with account
of NLO correction. See the text for more explanations of the behaviour
of the curves. Right: expected value of the variable $R_{\rho/\pi}\left(f_{\pi}/f_{\rho}\right)^{2}$
as a function of possible values of $a_{2}^{(\rho-\pi)}$ and $a_{4}^{(\rho-\pi)}$
for $x_{B}=0.1$ and $Q^{2}=4\,{\rm GeV}^{2}$. For the case of asymptotic
form distributions of both mesons ($a_{2}^{(\rho-\pi)}=a_{4}^{(\rho-\pi)}=0$)
the variable $R_{\rho/\pi}\left(f_{\pi}/f_{\rho}\right)^{2}=1$.}
\end{figure}

In order to quantize the sensitivity of the cross-section to deviation
of the meson DA from its asymptotic form, in Figure~\ref{fig:DVMP-r2n}
we show the dependence of the first two coefficients $r_{2}\left(x_{B},\,Q^{2}\right)$
and $r_{4}\left(x_{B},\,Q^{2}\right)$, defined in~(\ref{eq:RRatio}),
as functions of $x_{B}$ and $Q^{2}$. These coefficients do not
depend on the energy of the electron beam $E$, because at fixed $\left(x_{B},\,Q^{2}\right)$
the dependence on $E$ contributes only via a common $y$-dependent
prefactor in~(\ref{eq:Pref}), which does not contribute to $r_{2n}$.
The dependence of $r_{2n}$ on $Q^{2}$ is very mild and is due to the
logarithmic dependence of running coupling in the NLO contribution.
The dependence of $r_{2n}$ on $x_{B}$ exists due to the different $x_{B}$-dependence
of the leading order and next-to-lading order amplitudes. The fact
that the evaluated ratios $r_{2n}$ have a very mild dependence on
$Q^{2}$ and on $x_{B}$ (for $x_{B}\lesssim0.3$) implies that the
ratio of the cross-sections~(\ref{eq:RRatio}) only mildly depends on
$(x_{B},\,Q^{2})$, and its value is almost entirely determined by
the values of parameters 
\begin{equation}
a_{2}=a_{2}^{(\rho)}-a_{2}^{(\pi)},\qquad a_{4}=a_{4}^{(\rho)}-a_{4}^{(\pi)}.
\end{equation}
As can be seen from the Figure~\ref{fig:DVMP-r2n}, for the currently
expected phenomenological values of parameters $a_{2},a_{4}$ in the
range (\ref{eq:a2n_Estimates}), the ratio~(\ref{eq:RRatio})
changes up to 20\%. Since the expected values of $a_{2},\,a_{4}$ are
quite small, we may neglect the contributions of quadratic terms, so we
expect that $R_{\rho/\pi}$ is mostly sensitive to the combination
\begin{equation}
\left(a_{2}+\frac{r_{4}\left(x_{B},\,Q^{2}\right)}{r_{2}\left(x_{B},\,Q^{2}\right)}a_{4}\right).
\end{equation}
Given that the functions $r_{2}\left(x,\,Q^{2}\right)$, $r_{4}\left(x,\,Q^{2}\right)$
are known, measurement of $R_{\rho/\pi}$ in a sufficiently large kinematical
range could allow us to extract separately the values of $a_{2}$
and $a_{4}$.

\begin{figure}
\includegraphics[height=6cm]{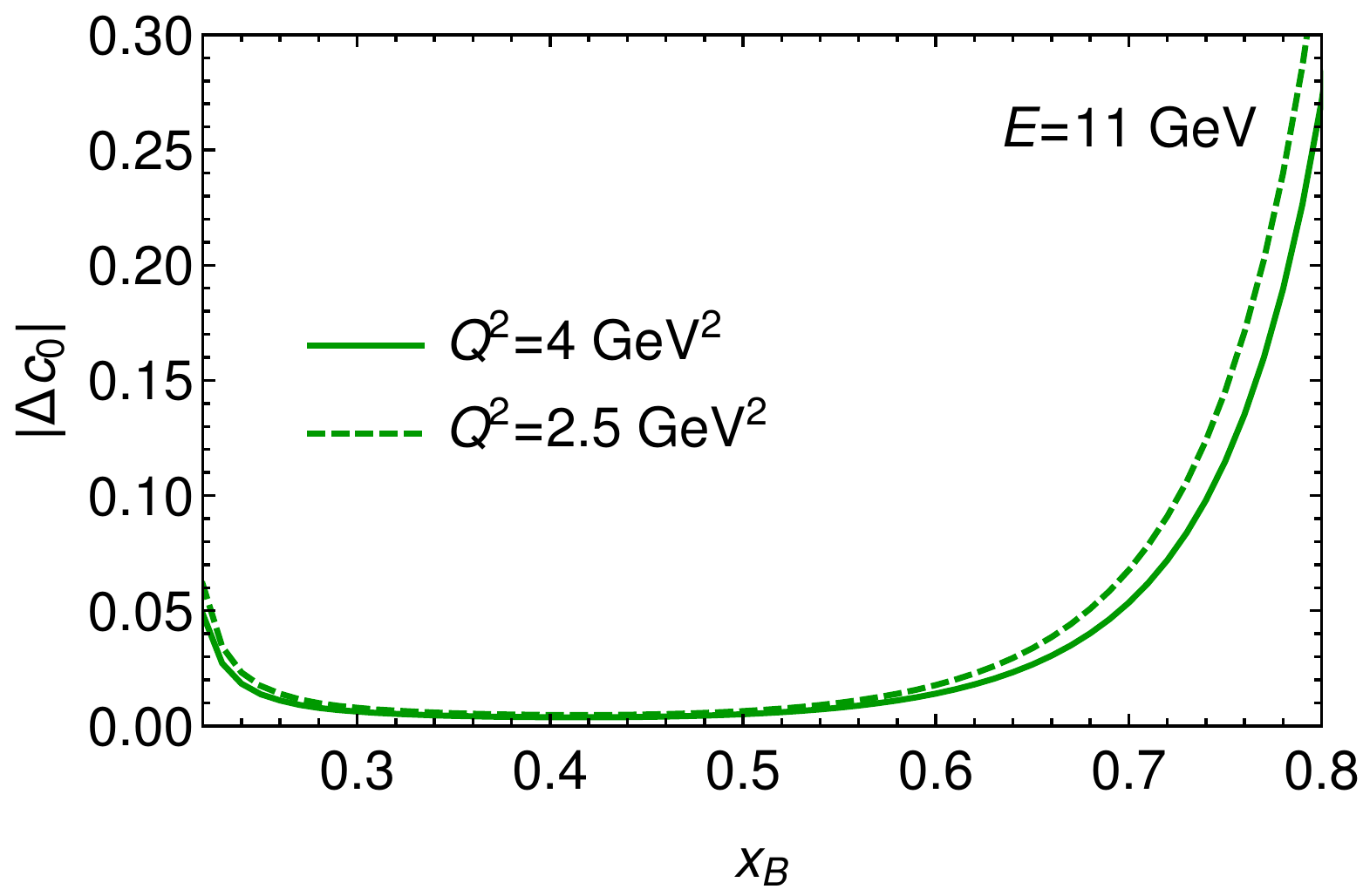}

\protect\caption{\label{fig:DVMP-harmonics}(color online) Upper values of the coefficient
$\left|\Delta c_{0}\right|$ for several values of $Q^{2}$, within JLab
kinematics ($E=11$~GeV).}
\end{figure}

As we explained in the previous section, for the case of the twist-three
harmonics, we are only interested in the contribution of the term
$c_{0}$ in~(\ref{eq:Harmonics}), which is the only term contributing
to the $\varphi$-integrated cross-sections. From Figure~\ref{fig:DVMP-harmonics},
we can see that the contribution of this term in the region of interest
is negligible and does not exceed a few per cent. Its relative contribution
increases in the region $x_{B}\gtrsim0.6-0.7$ and it might reach up
to 10 per cent. However, the cross-section is strongly suppressed
in that region, and the experimental statistics is quite poor, so for
this reason we expect that this region will not give a strong constraint
on the constructed parameterizations of the GPDs. In the region $x_{B}\approx0.1-0.3$,
which gives the dominant contribution within JLab kinematics, we expect
that the effects of the higher twist corrections will give just a
couple of per cent correction, and will not affect significantly the
ratio $R\left(a_{2},\,a_{4}\right)$, shown in the right panel of
Figure~\ref{fig:DVMP-r2n}. The effect of higher twist corrections
decreases as a function of $Q$ and becomes almost negligible for
$Q^{2}\gtrsim10\,{\rm GeV}^{2}.$

For deeply virtual meson production in other channels (\emph{e.g}.
production of kaons and $K^{*}$-mesons) the cross-sections have a similar
shape, although their values are smaller. Besides, the amplitudes
of these processes get comparable contributions from GPDs of different
partons, and for this reason the restrictions imposed by experimental
data on GPDs of individual partons are less binding~(see~\cite{Siddikov:2017nku}
for more details). Moreover, experimentally these channels present
more challenges and therefore will not
be considered here. The contribution of the higher twist corrections
might be estimated similarly in terms of higher twist harmonics.

\section{Conclusions}

In this paper we studied the contributions for $\rho$-meson production
in Bjorken kinematics. We found that the production
of both parity conjugate mesons ($\rho$ and $\pi$) in charged current
processes allows for a very clean probe of the generalized parton
distributions, and the ratio~(\ref{eq:RRatio}) provides the possibility
of clearly distinguishing contributions of higher twist corrections.
More precisely, since the cross-sections of both processes are sensitive
to the same set of GPDs, the ratio~(\ref{eq:RRatio}) should be almost
constant in the case of the leading twist dominance, and the value of
this constant depends only on the DAs of the produced mesons. The
presence of large higher twist corrections would reveal itself via
a pronounced dependence of the ratio~(\ref{eq:RRatio}) on both $x_{B}$
and $Q^{2}$. We expect that such processes might be studied either
in JLab future neutrino-induced experiments or in electron-induced
experiments in JLab and EIC. We estimated the cross-sections in the
kinematics of upgraded 12 GeV Jefferson Laboratory experiments, as
well as in the kinematics of the future Electron Ion Collider, and
found that the process can be measured with reasonable statistics.
A code for the evaluation of the cross-sections with various GPD models
is available on demand.

\section*{Acknowledgments}

This research was partially supported by Proyecto Basal FB 0821 (Chile),
the Fondecyt (Chile) grant1180232 and\lyxdeleted{Linux User}{Mon Apr  1 12:54:54 2019}{
} CONICYT (Chile) grant PIA ACT1413. Powered@NLHPC: This research
was partially supported by the supercomputing infrastructure of the
NLHPC (ECM-02). Also, we thank Yuri Ivanov for technical support of
the USM HPC cluster where part of evaluations were done.

\appendix

\section{NLO coefficient function}

\label{sec:NLOCoef}The function $T^{(1)}\left(v,\,z\right)$ in~(\ref{eq:c2-1})
encodes NLO corrections to the coefficient function. Explicitly, this
function is given by 
\begin{align}
T^{(1)}\left(v,\,z\right) & =\frac{1}{2vz}\left[\frac{4}{3}\left([3+\ln(v\,z)]\,\ln\left(\frac{Q^{2}}{\mu_{F}^{2}}\right)+\frac{1}{2}\ln^{2}\left(v\,z\right)+3\ln(v\,z)-\frac{\ln\bar{v}}{2\bar{v}}-\frac{\ln\bar{z}}{2\bar{z}}-\frac{14}{3}\right)\right.\label{eq:T1}\\
 & +\beta_{0}\left(\frac{5}{3}-\ln(v\,z)-\ln\left(\frac{Q^{2}}{\mu_{R}^{2}}\right)\right)\nonumber \\
 & -\frac{1}{6}\left(2\frac{\bar{v}\,v^{2}+\bar{z}\,z^{2}}{(v-z)^{3}}\left[{\rm Li}_{2}(\bar{z})-{\rm Li}_{2}(\bar{v})+{\rm Li}_{2}(v)-{\rm Li}_{2}(z)+\ln\bar{v}\,\ln z-\ln\bar{z}\,\ln v\right]\right.\nonumber \\
 & +2\frac{v+z-2v\,z}{(v-z)^{2}}\ln\left(\bar{v}\bar{z}\right)+2\left[{\rm Li}_{2}(\bar{z})+{\rm Li}_{2}(\bar{v})-{\rm Li}_{2}(z)-{\rm Li}_{2}(v)+\ln\bar{v}\,\ln z+\ln\bar{z}\,\ln v\right]\nonumber \\
 & +\left.\left.4\frac{v\,z\,\ln(v\,z)}{(v-z)^{2}}-4\ln\bar{v}\,\ln\bar{z}-\frac{20}{3}\right)\right],\nonumber 
\end{align}
where $\beta_{0}=\frac{11}{3}N_{c}-\frac{2}{3}N_{f}$, ${\rm Li}_{2}(z)$
is the dilogarithm function, and $\mu_{R}$ and $\mu_{F}$ are the
renormalization and factorization scales respectively. For the vector
meson production in processes when the internal state of the hadron
is not changed, the additional contribution comes from gluons and
singlet (sea) quarks~\cite{Belitsky:2001nq,Ivanov:2004zv,Diehl:2007hd}~\footnote{For the sake of simplicity, we follow~\cite{Diehl:2007hd} and assume
that the factorization scale $\mu_{F}$ is the same for both the generalized
parton distribution and the pion distribution amplitude.}, 
\begin{eqnarray}
c^{(g)}\left(x,\xi\right) & = & \left(\int dz\frac{\phi_{2,\pi}\left(z\right)}{z\,(1-z)}\right)\frac{2\pi i}{3}\frac{\alpha_{s}\left(\mu_{R}^{2}\right)f_{M}}{Q}\frac{\xi}{\left(\xi+x-i0\right)\left(\xi-x-i0\right)}\left(1+\frac{\alpha_{s}\left(\mu_{r}^{2}\right)}{4\pi}\mathcal{I}^{(g)}\left(\frac{\xi-x}{2\xi},\,z\right)\right),\label{eq:c2-2}\\
\mathcal{I}^{(g)}\left(v,\,z\right) & = & \left(\ln\left(\frac{Q^{2}}{\mu_{F}^{2}}\right)-1\right)\left[\frac{\beta_{0}}{2}+C_{A}\left[\left(1-v\right)^{2}+v^{2}\right]\left(\frac{\ln\left(1-v\right)}{v}+\frac{\ln v}{1-v}\right)-\frac{C_{F}}{2}\left(\frac{v\ln v}{1-v}+\frac{\left(1-v\right)\ln\left(1-v\right)}{v}\right)\right.\nonumber \\
 &  & +\left.C_{F}\left(\frac{3}{2}+2\,z\ln\left(1-z\right)\right)\right]-2\,C_{F}-\frac{\beta_{0}}{2}\left(\ln\left(\frac{Q^{2}}{\mu_{R}^{2}}\right)-1\right)-\frac{C_{F}\left(1-2\,v\right)}{2\,\left(z-v\right)}R\left(z,\,v\right)\nonumber \\
 &  & +\frac{\left(2C_{A}-C_{F}\right)}{4}\left(\frac{v\,\ln^{2}v}{1-v}+\frac{\left(1-v\right)\ln^{2}\left(1-v\right)}{v}\right)+C_{F}(1+3\,z)\ln\left(1-z\right)+\nonumber \\
 &  & +\left(\ln\,v+\ln\left(1-v\right)\right)\left[C_{F}\left(1-z\right)\ln\,z-\frac{1}{4}+2C_{F}-C_{A}\right]\nonumber \\
 &  & +\frac{C_{A}}{2}\left(\ln\left(z\,(1-z)\right)-2\right)\left[\frac{v\,\ln v}{1-v}+\frac{\left(1-v\right)\,\ln\left(1-v\right)}{v}\right]\label{eq:T1-a}\\
 &  & +C_{F}z\,\ln^{2}\left(1-z\right)+\frac{C_{A}}{2}\left(1-2\,v\right)\ln\left(\frac{v}{1-v}\right)\left[\frac{3}{2}+\ln\left(z\,\left(1-z\right)\right)+\ln\left(v\,\left(1-v\right)\right)\right]\nonumber \\
 &  & +\left(C_{F}\left((z-v)^{2}-v\,(1-v)\right)-\left(C_{F}-\frac{C_{A}}{2}\right)(z-v)(1-2v)\right)\times\nonumber \\
 &  & \times\left[-\frac{R(z,v)}{(z-v)^{2}}+\frac{\ln v+\ln z-\ln\left(1-v\right)-\ln\left(1-z\right)}{2\left(z-v\right)}+\frac{(z-v)^{2}-v(1-v)}{(z-v)^{3}}H(z,v)\right]\nonumber \\
 &  & +\left\{ \frac{}{}z\to1-z\right\} ,\nonumber \\
C_{F} & = & \frac{N_{c}^{2}-1}{2N_{c}},\quad C_{A}=N_{c}.
\end{eqnarray}
\begin{eqnarray}
c_{\pm}^{(s)}\left(x,\xi\right) & = & -\left(\int dz\frac{\phi_{2,\pi}\left(z\right)}{z\,(1-z)}\right)\frac{4i\alpha_{s}^{2}\left(\mu_{R}^{2}\right)f_{M}}{9\,Q}\mathcal{I}^{(s)}\left(\frac{x\pm\xi}{2\xi},\,z\right),\label{eq:c2-1}\\
\mathcal{I}^{(s)}\left(v,\,z\right) & = & \left(1-2\,v\right)\left(\frac{\ln v}{1-v}+\frac{\ln(1-v)}{v}\right)\ln\left(\frac{Q^{2}z}{\mu_{F}^{2}}\right)+\frac{1-2v}{2}\left[\frac{\ln^{2}v}{1-v}+\frac{\ln^{2}\left(1-v\right)}{v}\right]\label{eq:T1-b}\\
 &  & -\frac{R(v,\,z)}{z-v}-\frac{\left(1-v\right)\ln\left(1-v\right)-v\ln v}{v\left(1-v\right)}+\frac{\left(z-v\right)^{2}-v\left(1-v\right)}{\left(z-v\right)^{2}}H\left(v,z\right)+\left\{ \frac{}{}z\to1-z\right\} ,\nonumber \\
R\left(v,\,z\right) & = & z\ln v+(1-z)\,\ln\left(1-v\right)+z\,\ln z+\left(1-v\right)\ln\left(1-v\right),\\
H\left(v,\,z\right) & = & {\rm Li}_{2}\left(1-v\right)-{\rm Li}_{2}\left(v\right)+{\rm Li}_{2}\left(z\right)-{\rm Li}_{2}\left(1-z\right)+\ln v\,\ln\left(1-z\right)-\ln\left(1-v\right)\ln\,z.
\end{eqnarray}

Some coefficient functions have non-analytic behavior $\sim\ln^{2}v$
for small $v\approx0$ ($x=\pm\xi\mp i0$), which signals that the
collinear approximation might be not valid near this point. This singularity
in the collinear limit occurs due to the omission of the small transverse
momentum $l_{M,\perp}$ of the quark inside a meson~\cite{Goloskokov:2009ia}.
For this reason the contribution of the region $|v|\sim l_{M,\perp}^{2}/Q^{2}$
for finite $Q^{2}$ (below the Bjorken limit) should be treated with
due care. However, a full evaluation of $T^{(1)}\left(v,\,z\right)$
beyond the collinear approximation (taking into account all higher
twist corrections) presents a challenging problem and has not been
done so far. It was observed in~\cite{Diehl:2007hd}, that the singular
terms might be eliminated by a redefinition of the renormalization
scale $\mu_{R}$, however near the point $v\approx0$ the scale $\mu_{R}^{2}$
becomes soft, $\mu_{R}^{2}\sim z\,v\,Q^{2}\lesssim l_{\perp}^{2}$
which is another manifestation that nonperturbative effects become
relevant. For this reason, sufficiently large value of $Q^{2}$ should
be used to mitigate contributions of higher twist effects. As we will
see below, for $Q^{2}\approx4$ GeV$^{2}$ the contribution of this
soft region is small, so the collinear factorization is reliable.

As was discussed in Section~(\ref{subsec:DAs}), the distribution
amplitudes might be represented as~(\ref{eq:conformalExpansion}),
with major contribution from the terms with $n=0,\,1$ and $2.$ The
corresponding expressions for the parton amplitudes~(\ref{eq:c2},\ref{eq:c2-2},\ref{eq:c2-1})
take a form~\cite{Diehl:2007hd}
\begin{eqnarray}
c_{\pm,n}^{(q)}\left(x,\xi\right) & = & \frac{8\pi i}{9}\frac{\alpha_{s}\left(\mu_{R}^{2}\right)f_{M}}{Q}\frac{3}{x\pm\xi\mp i0}\left(1+\frac{\alpha_{s}\left(\mu_{r}^{2}\right)}{4\pi}t_{a,\,n}\left(\frac{\xi\pm x}{2\xi}\right)\right),\label{eq:c2-3}
\end{eqnarray}
\begin{align}
t_{a,0}(y) & =\beta_{0}\left[\frac{19}{6}-\ln y-\ln\left(\frac{Q^{2}}{\mu_{R}^{2}}\right)\right]+C_{F}\biggl[\left(3+2\ln y\right)\,\ln\left(\frac{Q^{2}}{\mu_{GPD}^{2}}\right)-\frac{77}{6}-\left(\frac{1}{\bar{y}}-3\right)\ln y+\ln^{2}y\biggr]\nonumber \\
 & \quad+\left(2C_{F}-C_{A}\right)\biggl\{-\frac{1}{3}-4(2-3y)\ln\bar{y}+2(1-6y)\ln y+4(1-3y)\,\bigl({\rm Li}_{2}y-{\rm Li}_{2}\bar{y}\bigr)\nonumber \\
 & \qquad\qquad+2(1-6y\bar{y})\left[3\bigl({\rm Li}_{3}\bar{y}+{\rm Li}_{3}y\bigr)-\ln y\,{\rm Li}_{2}y-\ln\bar{y}\,{\rm Li}_{2}\bar{y}-\frac{\pi^{2}}{6}\,\bigl(\ln y+\ln\bar{y}\bigr)\right]\biggr\}\,.
\end{align}
\begin{align}
t_{a,2}(y) & =\beta_{0}\left[\frac{21}{4}-\ln y-\ln\frac{Q^{2}}{\mu_{R}^{2}}\right]+C_{F}\biggl[\left(3+2\ln y\right)\,\ln\frac{Q^{2}}{\mu_{GPD}^{2}}-\frac{25}{6}\,\ln\frac{Q^{2}}{\mu_{DA}^{2}}-\frac{1019}{72}-\left(\frac{1}{\bar{y}}+\frac{7}{6}\right)\ln y+\ln^{2}y\biggr]\nonumber \\
 & \quad+\left(2C_{F}-C_{A}\right)\biggl\{\frac{401}{12}-255y+270y^{2}-\left(\frac{299}{3}-867y+1830y^{2}-1080y^{3}\right)\ln\bar{y}\nonumber \\
 & \qquad+\left(\frac{56}{3}-357y+1290y^{2}-1080y^{3}\right)\ln y+2\,\bigl(22-291y+780y^{2}-540y^{3}\bigr)\,\bigl({\rm Li}_{2}y-{\rm Li}_{2}\bar{y}\bigr)\nonumber \\
 & \qquad+12\,\left(1-21y+106y^{2}-175y^{3}+90y^{4}\right)\nonumber \\
 & \qquad\quad\times\biggl[3\bigl({\rm Li}_{3}\bar{y}+{\rm Li}_{3}y\bigr)-\ln y\,{\rm Li}_{2}y-\ln\bar{y}\,{\rm Li}_{2}\bar{y}-\frac{\pi^{2}}{6}\,\bigl(\ln y+\ln\bar{y}\bigr)\biggr]\biggr\}\,,\nonumber \\
t_{a,4}(y) & =\beta_{0}\left[\frac{31}{5}-\ln y-\ln\frac{Q^{2}}{\mu_{R}^{2}}\right]\nonumber \\
 & \quad+C_{F}\biggl[\left(3+2\ln y\right)\,\ln\frac{Q^{2}}{\mu_{GPD}^{2}}-\frac{91}{15}\,\ln\frac{Q^{2}}{\mu_{DA}^{2}}-\frac{10213}{900}-\left(\frac{1}{\bar{y}}+\frac{46}{15}\right)\ln y+\ln^{2}y\biggr]\nonumber \\
 & \quad+\left(2C_{F}-C_{A}\right)\biggl\{\frac{4903}{40}-\frac{5775}{2}y+\frac{57085}{4}y^{2}-23310y^{3}+11970y^{4}\nonumber \\
 & \qquad-\left(\frac{21109}{60}-\frac{41451}{5}y+\frac{103285}{2}y^{2}-125020y^{3}+129150y^{4}-47880y^{5}\right)\ln\bar{y}\nonumber \\
 & \qquad+\left(\frac{2899}{60}-\frac{11001}{5}y+\frac{45535}{2}y^{2}-78400y^{3}+105210y^{4}-47880y^{5}\right)\ln y\nonumber \\
 & \qquad+\bigl(137-4506y+35280y^{2}-100380y^{3}+117180y^{4}-47880y^{5}\bigr)\,\bigl({\rm Li}_{2}y-{\rm Li}_{2}\bar{y}\bigr)\phantom{\biggl[\biggr]}\nonumber \\
 & \qquad+30\,\bigl(1-48y+580y^{2}-2590y^{3}+5166y^{4}-4704y^{5}+1596y^{6}\bigr)\phantom{\biggl[\biggr]}\nonumber \\
 & \qquad\quad\times\biggl[3\bigl({\rm Li}_{3}\bar{y}+{\rm Li}_{3}y\bigr)-\ln y\,{\rm Li}_{2}y-\ln\bar{y}\,{\rm Li}_{2}\bar{y}-\frac{\pi^{2}}{6}\,\bigl(\ln y+\ln\bar{y}\bigr)\biggr]\biggr\}\,.
\end{align}
\begin{eqnarray}
c_{n}^{(g)}\left(x,\xi\right) & = & \frac{2\pi i\alpha_{s}\left(\mu_{R}^{2}\right)f_{M}}{Q}\frac{\xi}{\left(\xi+x-i0\right)\left(\xi-x-i0\right)}\left(1+\frac{\alpha_{s}\left(\mu_{r}^{2}\right)}{4\pi}t_{g,\,n}\left(\frac{\xi-x}{2\xi}\right)\right),\label{eq:c2-2-1}
\end{eqnarray}
where
\begin{align*}
t_{g,0}(y) & =\biggl[2C_{A}\,(y^{2}+\bar{y}^{2})-C_{F}\,y\biggr]\frac{\ln y}{\bar{y}}\,\ln\frac{Q^{2}}{\mu_{GPD}^{2}}+\frac{\beta_{0}}{2}\,\ln\frac{\mu_{R}^{2}}{\mu_{GPD}^{2}}\\
 & \quad+C_{F}\biggl[-\frac{5}{2}+\left(\frac{1}{\bar{y}}+1-4y\right)\ln y-\frac{y}{2}\,\frac{\ln^{2}y}{\bar{y}}\\
 & \qquad\qquad-2(\bar{y}-y){\rm Li}_{2}\bar{y}-4y\bar{y}\biggl(3{\rm Li}_{3}\bar{y}-\ln y\,{\rm Li}_{2}y-\frac{\pi^{2}}{6}\ln y\biggr)\biggr]\\
 & \quad+C_{A}\left[-\left(\frac{6}{\bar{y}}-8y\right)\ln y+\left(\frac{1}{\bar{y}}-2y\right)\ln^{2}y+2(\bar{y}-y){\rm Li}_{2}\bar{y}\right]+\{y\to\bar{y}\}\,,
\end{align*}
\begin{align}
t_{g,2}(y) & =\biggl[2C_{A}\,(y^{2}+\bar{y}^{2})-C_{F}\,y\biggr]\frac{\ln y}{\bar{y}}\,\ln\frac{Q^{2}}{\mu_{GPD}^{2}}+\frac{\beta_{0}}{2}\,\ln\frac{\mu_{R}^{2}}{\mu_{GPD}^{2}}-\frac{25}{12}\,C_{F}\ln\frac{Q^{2}}{\mu_{DA}^{2}}\nonumber \\
 & \quad+C_{F}\biggl[\frac{35}{36}(5-54y\bar{y})-\frac{y}{2}\,\frac{\ln^{2}y}{\bar{y}}-7(\bar{y}-y)\,(1-30y\bar{y}){\rm Li}_{2}\bar{y}\nonumber \\
 & \qquad\qquad+\left(\frac{1}{\bar{y}}-\frac{3}{2}-\frac{392}{3}y+525y^{2}-420y^{3}\right)\ln y\biggr]\nonumber \\
 & \quad+C_{A}\biggl[-\frac{15}{4}\,(1-4y\bar{y})+\left(\frac{1}{\bar{y}}-2y\right)\ln^{2}y+(\bar{y}-y)\,(7-60y\bar{y}){\rm Li}_{2}\bar{y}\nonumber \\
 & \qquad\qquad-\left(\frac{23}{3\bar{y}}+\frac{5}{6}-58y+150y^{2}-120y^{3}\right)\ln y\biggr]\nonumber \\
 & \quad+6y\bar{y}\,\Bigl[5(1-4y\bar{y})\,C_{A}-14(1-5y\bar{y})\,C_{F}\Bigr]\left(3{\rm Li}_{3}\bar{y}-\ln y\,{\rm Li}_{2}y-\frac{\pi^{2}}{6}\ln y\right)+\{y\to\bar{y}\}\,,\nonumber \\
t_{g,4}(y) & =\biggl[2C_{A}\,(y^{2}+\bar{y}^{2})-C_{F}\,y\biggr]\frac{\ln y}{\bar{y}}\,\ln\frac{Q^{2}}{\mu_{GPD}^{2}}+\frac{\beta_{0}}{2}\,\ln\frac{\mu_{R}^{2}}{\mu_{GPD}^{2}}-\frac{91}{30}\,C_{F}\ln\frac{Q^{2}}{\mu_{DA}^{2}}\nonumber \\
 & \quad+C_{F}\biggl[\frac{27287}{1800}-595y\bar{y}+2520(y\bar{y})^{2}-\frac{y}{2}\,\frac{\ln^{2}y}{\bar{y}}+16(\bar{y}-y)\Bigl(1-105y\bar{y}+630(y\bar{y})^{2}\Bigr){\rm Li}_{2}\bar{y}\nonumber \\
 & \qquad\qquad+\left(\frac{1}{\bar{y}}-\frac{5}{2}-\frac{11596}{15}y+9660y^{2}-34160y^{3}+45360y^{4}-20160y^{5}\right)\ln y\,\biggr]\nonumber \\
 & \quad+C_{A}\biggl[-\frac{35}{16}\,(1-4y\bar{y})(5-72y\bar{y})+\left(\frac{1}{\bar{y}}-2y\right)\ln^{2}y+2(\bar{y}-y)\Bigl(8-315y\bar{y}+1260(y\bar{y})^{2}\Bigr){\rm Li}_{2}\bar{y}\nonumber \\
 & \qquad\qquad-\left(\frac{257}{30\bar{y}}+\frac{77}{60}-\frac{1741}{5}y+2940y^{2}-8960y^{3}+11340y^{4}-5040y^{5}\right)\ln y\biggr]\nonumber \\
 & \quad+30y\bar{y}\,\Bigl[7(1-4y\bar{y})(1-6y\bar{y})\,C_{A}\nonumber \\
 & \qquad\qquad-16\,\Bigl(1-14y\bar{y}+42(y\bar{y})^{2}\Bigr)\,C_{F}\Bigr]\left(3{\rm Li}_{3}\bar{y}-\ln y\,{\rm Li}_{2}y-\frac{\pi^{2}}{6}\ln y\right)+\{y\to\bar{y}\}.
\end{align}
 The corresponding coefficients $r_{2n}\left(x,\,Q^{2}\right)$ which
define the sensitivity to harmonics are given by the ratios of the
amplitudes evaluated with convolution of the amplitudes with corresponding
GPDs, are related to the amplitudes as
\begin{align}
\frac{d\sigma}{dt\,dx_{B}dQ^{2}} & =\Gamma\sum_{\nu\nu'}\mathcal{A}_{\nu',\nu L}^{*}\mathcal{A}_{\nu',\nu L},\label{eq:sigma_def-2}
\end{align}
\[
r_{2n}\left(x,\,Q^{2},\,t\right)=\frac{1}{2}\left.\frac{\partial\left(d\sigma/dt\,dx_{B}dQ^{2}\right)}{\partial a_{2n}}\right|_{a_{2n}=0}=\frac{{\rm Re}\left(\sum_{\nu\nu'}\mathcal{A}_{\nu',\nu L}^{*(0)}\mathcal{A}_{\nu',\nu L}^{(2n)}\right)}{\sum_{\nu\nu'}\mathcal{A}_{\nu',\nu L}^{*(0)}\mathcal{A}_{\nu',\nu L}^{(0)}}
\]
where the superscript $(0)$ in the amplitudes $\mathcal{A}$ stands
for evaluation with asymptotic distribution amplitude, and superscript
$(2n)$ correspond to evaluation with distribution amplitude given
only by the $n$th term in~(\ref{eq:conformalExpansion}).


\begin{thebibliography}{100}
 \bibitem{Ji:1998xh} X.~D.~Ji and J.~Osborne, Phys.\ Rev.\ D
 \textbf{58} (1998) 094018 {[}arXiv:hep-ph/9801260{]}.
 
 \bibitem{Collins:1998be} J.~C.~Collins and A.~Freund, Phys.\ Rev.\ D
 \textbf{59}, 074009 (1999).
 
 \bibitem{Dupre:2017hfs}R.~Dupré, M.~Guidal, S.~Niccolai and M.~Vanderhaeghen,
 arXiv:1704.07330 {[}hep-ph{]}.
 
 \bibitem{Mueller:1998fv} D.~Mueller, D.~Robaschik, B.~Geyer, F.~M.~Dittes
 and J.~Horejsi, Fortsch.\ Phys.\ \textbf{42}, 101 (1994) {[}arXiv:hep-ph/9812448{]}.
 
 \bibitem{Ji:1996nm} X.~D.~Ji, Phys.\ Rev.\ D \textbf{55}, 7114
 (1997).
 
 \bibitem{Ji:1998pc} X.~D.~Ji, J.\ Phys.\ G \textbf{24}, 1181
 (1998) {[}arXiv:hep-ph/9807358{]}.
 
 \bibitem{Radyushkin:1996nd} A.~V.~Radyushkin, Phys.\ Lett.\ B
 \textbf{380}, 417 (1996) {[}arXiv:hep-ph/9604317{]}.
 
 \bibitem{Radyushkin:1997ki} A.~V.~Radyushkin, Phys.\ Rev.\ D
 \textbf{56}, 5524 (1997).
 
 \bibitem{Radyushkin:2000uy} A.~V.~Radyushkin, arXiv:hep-ph/0101225.
 
 \bibitem{Collins:1996fb} J.~C.~Collins, L.~Frankfurt and M.~Strikman,
 Phys.\ Rev.\ D \textbf{56}, 2982 (1997).
 
 \bibitem{Brodsky:1994kf} S.~J.~Brodsky, L.~Frankfurt, J.~F.~Gunion,
 A.~H.~Mueller and M.~Strikman, Phys.\ Rev.\ D \textbf{50}, 3134
 (1994).
 
 \bibitem{Goeke:2001tz} K.~Goeke, M.~V.~Polyakov and M.~Vanderhaeghen,
 Prog.\ Part.\ Nucl.\ Phys.\ \textbf{47}, 401 (2001) {[}arXiv:hep-ph/0106012{]}.
 
 \bibitem{Diehl:2000xz} M.~Diehl, T.~Feldmann, R.~Jakob and P.~Kroll,
 Nucl.\ Phys.\ B \textbf{596}, 33 (2001) {[}Erratum-ibid.\ B \textbf{605},
 647 (2001){]} {[}arXiv:hep-ph/0009255{]}.
 
 \bibitem{Belitsky:2001ns} A.~V.~Belitsky, D.~Mueller and A.~Kirchner,
 Nucl.\ Phys.\ B \textbf{629}, 323 (2002) {[}arXiv:hep-ph/0112108{]}.
 
 \bibitem{Diehl:2003ny} M.~Diehl, Phys.\ Rept.\ \textbf{388}, 41
 (2003) {[}arXiv:hep-ph/0307382{]}.
 
 \bibitem{Belitsky:2005qn} A.~V.~Belitsky and A.~V.~Radyushkin,
 Phys.\ Rept.\ \textbf{418}, 1 (2005) {[}arXiv:hep-ph/0504030{]}.
 
 \bibitem{Kubarovsky:2011zz}V.~Kubarovsky {[}CLAS Collaboration{]},
 Nucl.~Phys.~Proc.~Suppl.~\textbf{219-220}, 118 (2011).
 
 \bibitem{Ahmad:2008hp} S.~Ahmad, G.~R.~Goldstein and S.~Liuti,
 Phys.\ Rev.\ D \textbf{79} (2009) 054014 {[}arXiv:0805.3568 {[}hep-ph{]}{]}.
 
 \bibitem{Goloskokov:2009ia}S.~V.~Goloskokov and P.~Kroll, Eur.~Phys.~J.~C
 \textbf{65}, 137 (2010) {[}arXiv:0906.0460 {[}hep-ph{]}{]}.
 
 \bibitem{Goloskokov:2011rd}S.~V.~Goloskokov and P.~Kroll, Eur.~Phys.~J.~A
 \textbf{47}, 112 (2011) {[}arXiv:1106.4897 {[}hep-ph{]}{]}.
 
 \bibitem{Goldstein:2012az}G.~R.~Goldstein, J.~O.~G.~Hernandez
 and S.~Liuti, arXiv:1201.6088 {[}hep-ph{]}.
 
 \bibitem{Anikin:2009bf}I.~V.~Anikin, D.~Y.~Ivanov, B.~Pire,
 L.~Szymanowski and S.~Wallon, Nucl.~Phys.~B \textbf{828}, 1 (2010)
 {[}arXiv:0909.4090 {[}hep-ph{]}{]}.
 
 \bibitem{Diehl:1998pd}M.~Diehl, T.~Gousset and B.~Pire, Phys.~Rev.~D
 \textbf{59}, 034023 (1999) {[}hep-ph/9808479{]}.
 
 \bibitem{Mankiewicz:1998kg}L.~Mankiewicz, G.~Piller and A.~Radyushkin,
 \textbf{10}, 307 (1999) {[}hep-ph/9812467{]}.
 
 \bibitem{Mankiewicz:1999tt}L.~Mankiewicz and G.~Piller, Phys.~Rev.~D
 \textbf{61}, 074013 (2000) {[}hep-ph/9905287{]}.
 
 \bibitem{Boussarie:2017umz}R.~Boussarie, B.~Pire, L.~Szymanowski
 and S.~Wallon, arXiv:1708.09164 {[}hep-ph{]}.
 
 \bibitem{Berger:2001xd}E.~R.~Berger, M.~Diehl and B.~Pire, Eur.~Phys.~J.~C
 \textbf{23}, 675 (2002) {[}hep-ph/0110062{]}.
 
 \bibitem{Pire:2008ea}B.~Pire, L.~Szymanowski and J.~Wagner, Phys.~Rev.~D
 \textbf{79}, 014010 (2009) {[}arXiv:0811.0321 {[}hep-ph{]}{]}.
 
 \bibitem{Boer:2015fwa}M.~Boër, M.~Guidal and M.~Vanderhaeghen,
 Eur.~Phys.~J.~A \textbf{51}, no. 8, 103 (2015).
 
 \bibitem{Muller:2012yq}D.~Mueller, B.~Pire, L.~Szymanowski and
 J.~Wagner, Phys.~Rev.~D \textbf{86}, 031502 (2012) {[}arXiv:1203.4392
 {[}hep-ph{]}{]}.
 
 \bibitem{Sawada:2016mao}T.~Sawada, W.~C.~Chang, S.~Kumano, J.~C.~Peng,
 S.~Sawada and K.~Tanaka, Phys.~Rev.~D \textbf{93}, no. 11, 114034
 (2016) {[}arXiv:1605.00364 {[}nucl-ex{]}{]}.
 
 \bibitem{Ivanov:2004vd}D.~Y.~Ivanov, A.~Schafer, L.~Szymanowski
 and G.~Krasnikov, Eur.~Phys.~J.~C \textbf{34}, no. 3, 297 (2004)
 Erratum: {[}Eur.~Phys.~J.~C \textbf{75}, no. 2, 75 (2015){]} {[}hep-ph/0401131{]}.
 
 \bibitem{Ivanov:2015hca}D.~Y.~Ivanov, B.~Pire, L.~Szymanowski
 and J.~Wagner, arXiv:1510.06710 {[}hep-ph{]}.
 
 \bibitem{Kofler:2014yka}S.~Kofler, P.~Kroll and W.~Schweiger,
 Phys.~Rev.~D \textbf{91}, 054027 (2015) {[}arXiv:1412.5367 {[}hep-ph{]}{]}.
 
 \bibitem{Accardi:2012qut}A.~Accardi \emph{et al}., Eur.~Phys.~J.~A
 \textbf{52}, no. 9, 268 (2016) {[}arXiv:1212.1701 {[}nucl-ex{]}{]}.
 
 \bibitem{Gautheron:2010wva}F.~Gautheron \emph{et al}. {[}COMPASS
 Collaboration{]}, SPSC-P-340, CERN-SPSC-2010-014.
 
 \bibitem{Kouznetsov:2016vvo}O.~Kouznetsov {[}COMPASS Collaboration{]},
 Nucl.~Part.~Phys.~Proc.~\textbf{270-272}, 36 (2016).
 
 \bibitem{Ferrero:2012ega}A.~Ferrero {[}COMPASS Collaboration{]},
 AIP Conf.~Proc.~\textbf{1523}, 75 (2012).
 
 \bibitem{Sandacz:2016kwh}A.~Sandacz {[}COMPASS Collaboration{]},
 J.~Phys.~Conf.~Ser.~\textbf{678}, no. 1, 012045 (2016).
 
 \bibitem{Sandacz:2017ctv}A.~Sandacz {[}COMPASS Collaboration{]},
 PoS QCDEV \textbf{2016}, 018 (2017).
 
 \bibitem{Silva:2013dta}L.~Silva, Few Body Syst.~\textbf{54}, no.
 7-10, 1075 (2013).
 
 \bibitem{Kroll:2016kvd}P.~Kroll, JPS Conf.~Proc.~\textbf{13},
 010014 (2017).
 
 \bibitem{Kroll:2019wug}P.~Kroll, arXiv:1901.11380 {[}hep-ph{]}.
 
 \bibitem{Kroll:2018uvl}P.~Kroll and K.~Passek-Kumeri\v{c}ki, Phys.~Rev.~D
 \textbf{97}, no. 7, 074023 (2018) {[}arXiv:1802.06597 {[}hep-ph{]}{]}.
 
 \bibitem{Anikin:2017fwu}I.~V.~Anikin \emph{et al}., Acta Phys.~Polon.~B
 \textbf{49}, 741 (2018) {[}arXiv:1712.04198 {[}nucl-ex{]}{]}.
 
 \bibitem{Kroll:2017hym}P.~Kroll, Eur.~Phys.~J.~A \textbf{53},
 no. 6, 130 (2017) {[}arXiv:1703.05000 {[}hep-ph{]}{]}.
 
 \bibitem{Airapetian:2017vit}A.~Airapetian \emph{et al}. {[}HERMES
 Collaboration{]}, Eur.~Phys.~J.~C \textbf{77}, no. 6, 378 (2017)
 {[}arXiv:1702.00345 {[}hep-ex{]}{]}.
 
 \bibitem{Kroll:2016aop}P.~Kroll, Few Body Syst.~\textbf{57}, no.
 11, 1041 (2016) {[}arXiv:1602.03803 {[}hep-ph{]}{]}.
 
 \bibitem{Favart:2015umi}L.~Favart, M.~Guidal, T.~Horn and P.~Kroll,
 Eur.~Phys.~J.~A \textbf{52}, no. 6, 158 (2016) {[}arXiv:1511.04535
 {[}hep-ph{]}{]}.
 
 \bibitem{Kumericki:2017gdc}K.~Kumeri\v{c}ki and D.~Müller, AIP
 Conf.~Proc.~\textbf{1819}, no. 1, 050004 (2017).
 
 \bibitem{Kumericki:2016ela}K.~Kumeri\v{c}ki and D.~Mueller, Int.~J.~Mod.~Phys.~Conf.~Ser.~\textbf{40},
 1660047 (2016).
 
 \bibitem{Duplancic:2018bum}G.~Duplan\v{c}i\'{c}, K.~Passek-Kumeri\v{c}ki,
 B.~Pire, L.~Szymanowski and S.~Wallon, JHEP \textbf{1811}, 179
 (2018) {[}arXiv:1809.08104 {[}hep-ph{]}{]}.
 
 \bibitem{Duplancic:2016bge}G.~Duplan\v{c}i\'{c}, D.~Müller and
 K.~Passek-Kumeri\v{c}ki, Phys.~Lett.~B \textbf{771}, 603 (2017)
 {[}arXiv:1612.01937 {[}hep-ph{]}{]}.
 
 \bibitem{Pire:2017yge}B.~Pire and L.~Szymanowski, Phys.~Rev.~D
 \textbf{96}, no. 11, 114008 (2017) {[}arXiv:1711.04608 {[}hep-ph{]}{]}.
 
 \bibitem{Defurne:2016eiy}M.~Defurne \emph{et al}., {[}Jefferson
 Lab Hall A Collaboration{]}, Phys.~Rev.~Lett.~\textbf{117}, no.
 26, 262001 (2016) {[}arXiv:1608.01003 {[}hep-ex{]}{]}.
 
 \bibitem{THorn}G. Huber \emph{et al}, ''Scaling Study of the L-T
 Separated Pion Electroproduction Cross Section at 11 GeV ``, \href{https://misportal.jlab.org/mis/physics/experiments/viewProposal.cfm?paperId=512}{JLAB experiment E12-07-105}, 
 
 \bibitem{Fu:2016yzx}H.~B.~Fu, X.~G.~Wu, W.~Cheng and T.~Zhong,
 Phys.~Rev.~D \textbf{94}, no. 7, 074004 (2016) {[}arXiv:1607.04937
 {[}hep-ph{]}{]}.
 
 \bibitem{Bali:2017gfr}G.~S.~Bali \emph{et al}., Eur.~Phys.~J.~C
 \textbf{78}, no. 3, 217 (2018) {[}arXiv:1709.04325 {[}hep-lat{]}{]}.
 
 \bibitem{Ivanov:2004zv}D.~Y.~Ivanov, L.~Szymanowski and G.~Krasnikov,
 JETP Lett.~\textbf{80}, 226 (2004) {[}Pisma Zh.~Eksp.~Teor.~Fiz.~\textbf{80},
 255 (2004){]} Erratum: {[}JETP Lett.~\textbf{101}, no. 12, 844 (2015){]}
 , {[}hep-ph/0407207{]}.
 
 \bibitem{Diehl:2007hd}M.~Diehl and W.~Kugler, Eur.~Phys.~J.~C
 \textbf{52}, 933 (2007) {[}arXiv:0708.1121 {[}hep-ph{]}{]}.
 
 \bibitem{Pire:2015iza}B.~Pire and L. Szymanowski, Phys.~Rev.~Lett.~\textbf{115}
 (2015), 092001 {[}arXiv:1505.00917 {[}hep-ph{]}{]}.
 
 \bibitem{Pire:2015vxa}B.~Pire and L.~Szymanowski, Acta Phys.~Polon.~Supp.~
 \textbf{8}, 883 (2015) {[}arXiv:1510.01869 {[}hep-ph{]}{]}.
 
 \bibitem{Pire:2016jtr}B.~Pire, L.~Szymanowski and J.~Wagner, EPJ
 Web Conf.~\textbf{112}, 01018 (2016) {[}arXiv:1601.07666 {[}hep-ph{]}{]}.
 
 \bibitem{Pire:2017lfj}B.~Pire, L.~Szymanowski and J.~Wagner, Phys.~Rev.~D
 \textbf{95}, no. 9, 094001 (2017) {[}arXiv:1702.00316 {[}hep-ph{]}{]}.
 
 \bibitem{Pire:2017tvv}B.~Pire, L.~Szymanowski and J.~Wagner, Phys.~Rev.~D
 \textbf{95}, no. 11, 114029 (2017) {[}arXiv:1705.11088 {[}hep-ph{]}{]}.
 
 \bibitem{Siddikov:2016zmt}M.~Siddikov and I.~Schmidt, Phys.~Rev.~D
 \textbf{95}, no. 1, 013004 (2017) {[}arXiv:1611.07294 {[}hep-ph{]}{]}.
 
 \bibitem{Drakoulakos:2004gn}D.~Drakoulakos \emph{et al.} {[}Minerva
 Collaboration{]}, 
  hep-ex/0405002.
 
 \bibitem{Androic:2013rhu}D.~Androic \emph{et al}. {[}Qweak Collaboration{]},
 Phys.~Rev.~Lett.~\textbf{111} (2013) no.14, 141803 {[}arXiv:1307.5275
 {[}nucl-ex{]}{]}.
 
 \bibitem{Alcorn:2004sb}J.~Alcorn \emph{et al}., Nucl.~Instrum.~Meth.~A
 \textbf{522}, 294 (2004).
 
 \bibitem{Kopeliovich:2014pea}B.~Z.~Kopeliovich, I.~Schmidt and
 M.~Siddikov, Phys.~Rev.~D \textbf{89}, no. 5, 053001 (2014) {[}arXiv:1401.1547
 {[}hep-ph{]}{]}.
 
 \bibitem{Kopeliovich:2013ae} B.~Z.~Kopeliovich, I.~Schmidt and
 M.~Siddikov, Phys.\ Rev.\ D \textbf{87}, 033008 (2013) {[}arXiv:1301.7014
 {[}hep-ph{]}{]}.
 
 \bibitem{Ball:2006wn}P.~Ball, V.~M.~Braun and A.~Lenz, JHEP \textbf{0605}
 (2006) 004 {[}arXiv:hep-ph/0603063{]}.
 
 \bibitem{Kopeliovich:2011rv}B.~Z.~Kopeliovich, Iván Schmidt and
 M.~Siddikov, Nucl.~Phys.~A \textbf{918}, 41 (2013) {[}arXiv:1108.5654
 {[}hep-ph{]}{]}.
 
 \bibitem{Ball:1998sk}P.~Ball, V.~M.~Braun, Y.~Koike and K.~Tanaka,
 Nucl.~Phys.~B \textbf{529}, 323 (1998) {[}hep-ph/9802299{]}.
 
 \bibitem{Vanderhaeghen:1998uc}M.~Vanderhaeghen, P.~A.~M.~Guichon
 and M.~Guidal, 
  Phys.~Rev.~Lett.~\textbf{80}, 5064 (1998).
 
 \bibitem{Goloskokov:2006hr}S.~V.~Goloskokov and P.~Kroll, 
  Eur.~Phys.~J.~C \textbf{50}, 829 (2007) {[}hep-ph/0611290{]}.
 
 \bibitem{Goloskokov:2007nt}S.~V.~Goloskokov and P.~Kroll, 
  Eur.~Phys.~J.~C \textbf{53}, 367 (2008) {[}arXiv:0708.3569 {[}hep-ph{]}{]}.
 
 \bibitem{Goloskokov:2008ib}S.~V. Goloskokov and P.~Kroll, Eur.~Phys.~J.~C
 \textbf{59} (2009) 809 {[}arXiv:0809.4126 {[}hep-ph{]}{]}.
 
 \bibitem{Kopeliovich:2012dr}B.~Z.~Kopeliovich, I.~Schmidt and
 M.~Siddikov, Phys. Rev. D \textbf{86} (2012), 113018 {[}arXiv:1210.4825
 {[}hep-ph{]}{]}.
 
 \bibitem{Frankfurt:1999fp}L.~L.~Frankfurt, P.~V.~Pobylitsa, M.~V.~Polyakov
 and M.~Strikman, 
  Phys.~Rev.~D \textbf{60} (1999) 014010 {[}hep-ph/9901429{]}.
 
 \bibitem{Anikin:2009hk}I.~V.~Anikin, D.~Y.~Ivanov, B.~Pire,
 L.~Szymanowski and S.~Wallon, Phys.~Lett.~B \textbf{682}, 413
 (2010) {[}arXiv:0903.4797 {[}hep-ph{]}{]}.
 
 \bibitem{Siddikov:2017nku}M.~Siddikov and I.~Schmidt, Phys.~Rev.~D
 \textbf{96}, no. 9, 096006 (2017) {[}arXiv:1709.01405 {[}hep-ph{]}{]}.
 
 \bibitem{Belitsky:2001nq}A.~V.~Belitsky and D.~Mueller, Phys.~Lett.~B
 \textbf{513}, 349 (2001) {[}hep-ph/0105046{]}.
 
\end{thebibliography}
 \end{document}